\documentclass[10pt,aps,prd,twocolumn,showkeys,nofootinbib,floatfix,amsmath,amssymb,showpacs,superscriptaddress,notitlepage]{revtex4-1}

\usepackage{rotating}
\usepackage{graphicx}
\usepackage{dcolumn}
\usepackage{bm}
\usepackage{color}
\usepackage{multirow}
\usepackage{mathtools}
\usepackage{tabularx}
\newcolumntype{C}[1]{>{\hsize=#1\hsize\centering\arraybackslash}X}
\newcolumntype{R}[1]{>{\hsize=#1\hsize\raggedleft\arraybackslash}X}
\newcolumntype{L}[1]{>{\hsize=#1\hsize\raggedright\arraybackslash}X}
\usepackage{pifont}
\def\y{\ding{51}}
\def\n{\ding{55}}

\usepackage[caption=false]{subfig}
\usepackage[normalem]{ulem}  % \sout{old text} for strikeout
\usepackage{xr-hyper}
\usepackage[pdftex,bookmarks=true]{hyperref}
\hypersetup{
   colorlinks,
   citecolor=black,
   filecolor=black,
   linkcolor=black,
   urlcolor=black
}
\usepackage{cleveref}
\allowdisplaybreaks

\renewcommand\sout{\bgroup \color{red} \ULdepth=-.5ex \ULset}

\begin{document}

	\title{Charmed baryon spectrum from lattice QCD near the physical point}

	\author{H. Bahtiyar}
	\affiliation{Department of Physics, Mimar Sinan Fine Arts University, Bomonti 34380 Istanbul Turkey}
	\author{K. U. Can}
	\affiliation{CSSM, Department of Physics, The University of Adelaide, Adelaide SA 5005, Australia}
	\affiliation{RIKEN Nishina Center, RIKEN, Saitama 351-0198, Japan}
	\author{G. Erkol}
	\affiliation{Department of Natural and Mathematical Sciences, Faculty of Engineering, Ozyegin University, Nisantepe Mah. Orman Sok. No:34-36, Alemdag 34794 Cekmekoy, Istanbul Turkey}
	\author{P. Gubler}
	\affiliation{Advanced Science Research Center, Japan Atomic Energy Agency, Tokai, Ibaraki, 319-1195 Japan}
	\author{M. Oka}
	\affiliation{Advanced Science Research Center, Japan Atomic Energy Agency, Tokai, Ibaraki, 319-1195 Japan}
	\author{T. T. Takahashi}
	\affiliation{National Institute of Technology, Gunma College, Maebashi, Gunma 371-8530 Japan}

	\collaboration{TRJQCD Collaboration}
	\noaffiliation

	\date{\today}

	\begin{abstract}
		We calculate the low-lying spectrum of charmed baryons in lattice QCD on the $32^3\times64$, $N_f=2+1$ PACS-CS gauge configurations at the almost physical pion mass of $\sim 156$ MeV/c$^2$. By employing a set of interpolating operators with different Dirac structures and quark-field smearings for the variational analysis, we extract the ground and first few excited states of the spin-$1/2$ and spin-$3/2$, singly-, doubly-, and triply-charmed baryons. Additionally, we study the $\Xi_c$--$\Xi_c^\prime$ mixing and the operator dependence of the excited states in a variational approach. We identify several states that lie close to the experimentally observed excited states of the $\Sigma_c$, $\Xi_c$ and $\Omega_c$ baryons, including some of the $\Xi_c$ states recently reported by LHCb. Our results for the doubly- and triply-charmed baryons are suggestive for future experiments.
	\end{abstract}
	\maketitle

	\section{Introduction}
		Recent experimental results from the LHCb Collaboration on the $\Omega_c$ , $\Xi_c$ and the doubly charmed $\Xi_{cc}$ state have put further emphasis on the relevance of the hadron spectroscopy. There now exist 31 observed charmed baryons, 25 of which are classified with at least three stars by the Particle Data Group (PDG)~\cite{PhysRevD.98.030001}. Charmed baryons provide a unique laboratory to study the strong  interaction and confinement dynamics due to the composition of the light and charm quarks. Studying the excited states of the charmed baryons has the potential to reveal their internal dynamics and the nature of the excitation mechanisms.  

		Experimentally, the singly-charmed baryon sector is most accessible. Within this sector, the $\Lambda_c$ channel is most established. In addition to the ground state, there are four excitations with total spin up to $5/2$, although in need of a confirmation of the assigned quantum numbers. Out of the three $\Sigma_c$ states that are listed by the PDG, two are the lowest $J^P = 1/2^+$ and $3/2^+$ states and $\Sigma_c(2800)$ is their only observed excitation. This state has been detected in the $\Lambda_c \pi$ channel by the Belle~\cite{Mizuk:2004yu} and the BABAR~\cite{Aubert:2008ax} Collaborations. Its quantum numbers are not measured. In contrast, the $\Xi_c$ sector is quite rich since it can have flavor symmetric and antisymmetric wave functions. There are up to seven $\Xi_c$ excitations observed by the Belle~\cite{Li:2017uvv,Chistov:2006zj,Lesiak:2008wz,Yelton:2016fqw,Kato:2013ynr,Kato:2016hca}, the  BABAR~\cite{Aubert:2007eb,Aubert:2007dt,Aubert:2007dt} and very recently by the LHCb~\cite{Aaij:2020yyt} Collaborations in the energy range of $2920$ to $3120$ MeV/c$^2$. The PDG considers the existence of three of them to be very likely or certain while the confidence for the other two is smaller. LHCb states are not included in the review yet. These excited states appear in the invariant mass distributions of several singly-charmed baryon $B_c + \overline{K}$ or $\pi$ channels depending on the strangeness number of the baryon and in the $\Lambda D$ channel where the charm quark is confined in the meson system. This unique behavior makes the $\Xi_c$ system a good laboratory to study the internal excitation dynamics of the charmed baryons and the diquark correlations. The quantum numbers of these states remain undetermined. The LHCb Collaboration has also reported the precise measurements of the masses and the decay widths of five new $\Omega_c^0$ states \cite{PhysRevLett.118.182001}, which are observed in the $\Xi_c \overline{K}$ channel in the energy range from $3000$ to $3120$ MeV/c$^2$. Their spin-parity quantum numbers remain undetermined. There are several works in the literature investigating the nature of these states and assigning conflicting spin-parity quantum numbers. It is a triumph of the experiments to identify many states in such narrow energy windows. 

		The lowest-lying states of the singly-charmed baryons are already established by experimental studies and the lattice QCD results agree well with those observations. The $\Xi_{cc}$ is the only observed doubly-charmed baryon for the time being. It was first observed by the SELEX Collaboration~\cite{Mattson:2002vu,Ocherashvili:2004hi} but its results were not confirmed by other experiments until the LHCb Collaboration has reported the same particle with a different mass~\cite{PhysRevLett.119.112001}. Lattice QCD predictions for the mass of the $\Xi_{cc}$ lie above the SELEX reported value but agree very well with the LHCb value.  

		From the theoretical side, it remains to be a remarkable challenge to extract the spectrum and assign quantum numbers to the observed charmed baryons. 
		For a complete understanding of these states, one would in principle would need to study their decay widths as well. Spectra and properties of the heavy baryons have been studied extensively via several naive and improved quark models~\cite{PhysRevD.20.768,PhysRevD.34.2809,Silvestre:1996,PhysRevD.66.014008,PhysRevD.72.034026,EBERT2008612,PhysRevD.84.014025,Garcilazo_2007,Valcarce:2008ahj,Roberts:2007ni,Vijande:2013jath,Yoshida:2015tia,Shah_2016,Wang:2017hej,Kim:2017jpx,Karliner:2017kfm,Wang:2017vnc,Cheng:2017ove,Huang:2017dwn,Yang:2017rpg}, the Feynman-Hellmann theorem~\cite{PhysRevD.52.1722}, large N QCD~\cite{JENKINS1993447}, QCD sum rules~\cite{BAGAN1992367,BAGAN1992176,HUANG2000288,WANG201059,PhysRevD.91.054034,PhysRevD.65.094036,Wang:2008zg,PhysRevD.78.094007,PhysRevD.78.094015,Wang:2017zjw,Chen:2017sci,Agaev:2017jyt,Agaev:2020fut}, chiral effective-field theory~\cite{Wang:2020dhf}, chiral diquark effective theory~\cite{Harada:2019udr,Kim:2020imk} and heavy-quark effective theory~\cite{PhysRevD.56.R6738} approaches. Discussions about the excited $\Lambda_c$, $\Sigma_c$ and $\Xi_c$ states from various models are reviewed in detail in Refs.~\cite{Crede_2013,Chengs:2015hy}. Specifically, the excited $\Omega_c$ system is studied in the context of the QCD sum rules~\cite{Wang:2017zjw,Chen:2017sci,Agaev:2017jyt}, the constituent quark model~\cite{Wang:2017hej} and in a chiral quark-soliton model~\cite{Kim:2017jpx}. Calculations based on a quark-diquark bound state picture are presented in Refs.~\cite{Karliner:2017kfm,Wang:2017vnc,Cheng:2017ove} and arguments for a potential molecular~\cite{Huang:2017dwn}, or a compact pentaquark nature~\cite{Yang:2017rpg} for these states are given in other works. A dedicated lattice QCD study assigning quantum numbers is reported by the Hadron Spectrum Collaboration~\cite{PhysRevLett.119.042001}.        

		The lowest-lying charmed baryon states have been studied by various lattice groups as well. Early investigations utilized the quenched approximation~\cite{PhysRevD.54.3619,Flynn_2003,PhysRevD.66.014502,PhysRevD.64.094509,CHIU2005471}, while recent studies employ up to $2+1+1$-flavor dynamical gauge configurations with several lattice spacings, volumes and light-quark masses to estimate the baryon masses at the physical point~\cite{Alexandrou:2017xwd,Durr:2012dw,Brown:2014ena,Namekawa:2013vu,Chen:2017kxr,Briceno:2012wt,Alexandrou:2014sha,PhysRevD.92.034504,PhysRevD.99.031501,Padmanath:2013zfa,Padmanath:2015jea,Padmanath:2013bla,Padmanath:2014bxa,Padmanath:2015bra}. We summarize the recent studies of several lattice groups in \Cref{tab:lqcd_rw}.

		\begin{table*}[!htb]
			\caption{ \label{tab:lqcd_rw} Simulation properties of previous lattice QCD calculations. Works in the upper panel extract the ground states only while the ones in the lower panel study the excited states as well. We indicate the number of flavors ($N_f$), lattice spacing(s) ($a$), number of volumes ($n_V$) and the relevant sea- and valance-quark actions (S) used in the studies. Additionally, whether a relativistic treatment (RT) applied (\y) to the charm quark or not (\n) is indicated, and, in the last column, the chiral extrapolation method is quoted where applicable. NA (not applicable) means those groups run their simulations at the physical quark mass. Abbreviations are: highly-improved staggered quark (HISQ), relativistic heavy-quark action (RHQA), heavy-hadron chiral perturbation theory treatment (HH$\chi$PT), and Gell-Mann -- Okubo relation (GMO).} 
			\centering
			\setlength{\extrarowheight}{1pt}
			\begin{tabularx}{\textwidth}{L{1.2} C{1.0} C{.9} C{1.1} C{1} C{.5} C{1.2} C{1.2} C{.5} C{1} }
			\hline\hline
							 		& Ref.	& $N_f$ 	& $a$ [fm]		& $m_\pi$ [MeV] & $n_V$ & $S_{u,d,s,c}^{sea}$ 	& $S_{c}^{val}$ & RT & Extrapolation \\
				\hline
				ETM					& \cite{Alexandrou:2017xwd} 	
											& $2$ 		& 0.094			& 130 			& 1 	& Twisted Mass	 		& Twisted Mass	& \n & NA 			 \\
				D\"{u}r et al. 		& \cite{Durr:2012dw} 	
											& $2$		& 0.073			& 280			& 1		& Clover		 		& Brillouin 	& \y & \n 			 \\
				Brown et al. 		& \cite{Brown:2014ena} 	
											& $2+1$		& 0.085 - 0.11	& 227 - 419 	& 2 	& Domain Wall	 		& RHQA		   	& \y & HH$\chi$PT 	 \\
				PACS-CS 			& \cite{Namekawa:2013vu} 		
											& $2+1$		& 0.09			& 135			& 1		& Clover		 		& RHQA 	 		& \y & NA 			 \\
				TWQCD 				& \cite{Chen:2017kxr} 	
											& $2+1+1$ 	& 0.063			& 280			& 1 	& Domain Wall 	 		& Domain Wall 	& \n & \n 			 \\
				Brice\~{n}o et al. 	& \cite{Briceno:2012wt} 
											& $2+1+1$	& 0.06 - 0.12	& 220 - 310 	& 5 	& HISQ			 		& RHQA		 	& \y & HH$\chi$PT 	 \\
				ETM			& \cite{Alexandrou:2014sha}
											& $2+1+1$ 	& 0.094 - 0.065	& 210 - 430  	& 3 	& Twisted Mass	 		& Twisted Mass	& \n & HH$\chi$PT 	\\	
				\hline			
				RQCD 				& \cite{PhysRevD.92.034504} 		
											& $2+1$		& 0.075			& 259 - 460 	& 2		& Clover		 		& Clover 	 	& \y & GMO 			 \\
				HSC 				& \cite{Padmanath:2013zfa,Padmanath:2015jea,Padmanath:2013bla,Padmanath:2014bxa,Padmanath:2015bra} 
											& $2+1$		& 0.035			& 390 			& 1		& Clover		 		& Clover 	 	& \y & \n 			 \\
			\hline\hline
			\end{tabularx}
		\end{table*}

		There is a remarkable agreement between the results of the different groups utilizing different types of quark actions and approaches to the physical point. Most of those studies are motivated by the observation of the $\Xi_{cc}$ baryon by LHCb and thus their focus has been on the lowest-lying positive parity baryons. Extracting the excited states, however, is a challenge compared to calculating the ground states. The majority of the attention has been on the light-quark sector, especially on the Roper resonance and the $\Lambda(1405)$, while there are just a few groups that have studied the excited states of the charmed baryons.

		The RQCD Collaboration reported results for the singly- and doubly-charmed baryons, including excited states~\cite{PhysRevD.92.034504}. They employ several $2+1$-flavor gauge ensembles with a fixed lattice spacing but two different volumes and varying light-quark masses with the lightest one corresponding to a pion mass of $m_\pi \sim 260$ MeV/c$^2$. All the sea and valance quarks (including the charm quark) are treated via a non-perturbatively improved stout-smeared Clover action. The bare charm-quark mass is tuned to reproduce the $1S$ spin-averaged charmonium mass. In addition to spectrum calculations, they also investigate the light-flavor dependence of the singly and doubly charmed states. To this end, the operator set they use consists of interpolating fields based on $SU(4)$ symmetry and heavy quark effective theory (HQET) pictures. In order to access the excited states, they perform a variational analysis over a set of interpolating fields with three different quark-field smearings. Their chiral extrapolations follow a different approach compared to the other groups since they start from an $SU(3)$ symmetric point for the light and strange quarks and vary their masses while keeping the singlet quark mass fixed in their descent to the physical point. This leads to fits based on Gell-Mann -- Okubo relations for the charmed baryons. The lowest-lying extracted states are in good agreement with the other lattice determinations and with experimental values where available.     

		The Hadron Spectrum Collaboration (HSC) extracts the charmed baryon spectrum including positive and negative parity baryons with total spin up to $J=7/2$. They use $N_f=2+1$ anisotropic lattices generated with a tree-level tadpole-improved Clover fermion action with a pion mass of $m_\pi=391$ MeV/c$^2$. The anisotropic Clover action is used for the charm quark as well with its mass parameter tuned non-perturbatively so as to reproduce the dispersion relation for the $\eta_c$ meson. By using a large set of continuum interpolating operators, including \emph{nonlocal} covariant derivative operators, subduced to the irreducible representations of the cubic group, they form the basis for the variational correlation matrix analysis and extract the spectrum of the singly-, doubly- and triply-charmed baryons~\cite{Padmanath:2013zfa,Padmanath:2015jea,Padmanath:2013bla,Padmanath:2014bxa,Padmanath:2015bra}. Although the systematics are left unchecked and the pion mass is unphysical, their pioneering results provide valuable insight into the charmed baryon spectrum.

		In this work, we follow a conventional approach by using local operators only. Notable improvements of this study compared to the previous works that extract the excited baryon spectrum are the fully relativistic treatment of the charm quark in combination with the ``Clover'' action, thus the suppression of the $O(a m_Q)$ discretization errors, and working on gauge configurations with almost physical light quarks, hence eliminating the chiral extrapolation systematics. We also perform variational analyses over sets of operators with different Dirac structures and quark smearings and their combinations. Preliminary results of this work have been presented in Ref~\cite{Can:2019wts}. 

		This paper is structured as follows: we outline the approach to extract the baryon energies and the formulation of the variational analysis in \Cref{sec:lf}. Details of our lattice setup, the heavy quark action that we employ, and the choice of baryon operators are given in \Cref{sec:ls}. A detailed discussion on the variational analyses and the states we extract are presented in \Cref{sec:rd}. \Cref{sec:sc} holds the summary of our findings.   

	\section{Extracting excited states}\label{sec:lf}
		For a given interpolator, $\chi_i$, the two-point correlation function contains the contributions from all the states that couple to the corresponding quantum number,
		\begin{equation}\label{eq:twopt}
			C_{ij}(t) = \langle \chi_i(t) \bar{\chi}_j(0) \rangle = \sum_\mathcal{B} \langle 0 | \chi_i| \mathcal{B} \rangle \langle \mathcal{B} | \bar{\chi}_j| 0 \rangle e^{-E_\mathcal{B} t},
		\end{equation}
		where ($E_{\mathcal{B}}$) $\mathcal{B}$ stands for the (energy of the) baryon state. The desired parity state can be isolated by applying the parity operator, $P^{\pm} C_{ij}(t) = \frac{1}{2}(1 \pm \gamma_4) C_{ij}(t)$.

		Using a set of operators that couple to the same quantum numbers, one can utilize a variational approach to extract the tower of states. One can form an $N \times N$ correlation function matrix, 
		\begin{equation}
			C(t) = 
		 \begin{pmatrix}
		  C_{11}(t) & C_{12}(t) & \cdots \\
		  C_{21}(t) & C_{22}(t) & \cdots \\
		  \vdots  & \vdots  & \ddots
		 \end{pmatrix},
		\end{equation} 
		where each element, $C_{ij}(t)$, is an individual correlation function given in \Cref{eq:twopt}. Then, by solving the generalized eigenvalue problem~\cite{LUSCHER1990222,MICHAEL198558},
		\begin{align}\label{eq:gev}
			\begin{split}
				C(t)\psi^\alpha(t) &= \lambda^\alpha(t,t_0)C(t_0)\psi^\alpha(t), \\
				\phi^\alpha(t) C(t) &= \lambda^\alpha(t,t_0) \phi^\alpha(t) C(t_0),
			\end{split}
		\end{align}
		one extracts the left and right eigenvectors, $\psi^\alpha$ and $\phi^\alpha$, and uses them to diagonalize the correlation-function matrix,
		\begin{align}\label{eq:dc}
			\begin{split}
				\phi^\alpha(t^\prime) C(t) \psi^\beta(t^\prime) &\equiv C^\alpha(t) \\
				&= \delta_{\alpha \beta} Z^\alpha \bar{Z}^\beta e^{-E_\alpha t} \left( 1 + \mathcal{O}(e^{-\Delta E_\alpha t}) \right),  
			\end{split}
		\end{align}
		to access the energies of the states, $E_\alpha$. One can alternatively utilize the individual eigenvalues, $\lambda^\alpha(t,t_0) \sim e^{-E_\alpha (t-t_0)}(1+\mathcal{O}(e^{-\Delta E_{\alpha}t}))$, of the left and right eigenvalue equations given in \Cref{eq:gev} to extract the energies of the states. Both approaches give complementary results with some caveats\cite{PhysRevD.95.074507}. We prefer the method outlined above. Note that a suitable combination of the time-slice $t_0$ and the time slice of the eigenvectors, $t^\prime$, is chosen with respect to the quality and stability of the signal. Additionally, $t^\prime$ may or may not be chosen equal to $t$. Once the correlation function matrix is diagonalized, one can follow the standard techniques and perform an effective mass analysis for each state, $\alpha$,
		\begin{equation}
			m_{\text{eff}}^\alpha(t) = \text{ln} \frac{C^\alpha(t)}{C^\alpha(t+1)}.
		\end{equation}

	\section{Lattice Setup}\label{sec:ls}

		\subsection{Quark Actions}
			We employ the $32^3 \times 64$, $2+1$-flavor gauge configurations that are generated by the PACS-CS Collaboration~\cite{PhysRevD.79.034503}. These configurations are generated with the Iwasaki gauge action ($\beta=1.9$) and with the non-perturbatively $\mathcal{O}(a)$-improved Wilson (Clover) action ($c_{sw} = 1.715$) for the sea quarks. We perform our simulations on the $\kappa^{\text{sea}}_{ud}=0.13781$ subset, which have almost physical light quarks corresponding to $m_\pi = 156(9) \text{ MeV}/c^2$ as measured by PACS-CS. This subset has $m_\pi L = 2.3$, which would suggest sizable finite size effects. The hopping parameter of the strange quark is fixed to $\kappa^{\text{sea}}_{s} = 0.13640$. The scale is set via the masses of $\pi$, $K$, and $\Omega$ and the lattice spacing is determined to be $a=0.0907(13)$ fm ($a^{-1} = 2.176$ GeV).

			We use the Clover action for the valence $u/d$ and $s$ quarks. The hopping parameter of the valence light quarks is set equal to those of sea quarks, $\kappa^{\text{val}}_{u/d} = \kappa^{\text{sea}}_{u/d}$. Due to an overestimation of the mass of the $\Omega^-$ particle with $\kappa^{\text{val}}_{s} = \kappa^{\text{sea}}_{s}$, however, we re-tune the hopping parameter of the valence strange quark to $\kappa_s^{\text{val}} = 0.13665$, in order to match the physical $\Omega^-$ mass on these configurations. Details of this tuning are discussed in Ref.~\cite{PhysRevD.98.114505}.

			We employ a relativistic heavy-quark action for the charm quark,
			\begin{equation}
				S_\Psi = \sum_{x,y} \bar{\Psi}_x D_{x,y} \Psi_y,  	
			\end{equation}   
			where the $\Psi$s are the heavy quark spinors and the fermion matrix is given as
			\begin{align}
				\begin{split}
					D_{x,y} &= \delta_{xy} 	
					- \kappa_Q \sum_{\mu=1}^3 \left[ (r_s - \nu \gamma_\mu) U_{x,\mu} \delta_{x+\hat\mu,y} \right. \\
					& \left. + (r_s + \nu \gamma_\mu) U^\dag_{x,\mu} \delta_{x,y+\hat\mu} \right] - \kappa_Q \left[ (1 - \gamma_4) U_{x,4} \delta_{x+\hat4,y} \right. \\
					& \left. + (1 + \gamma_4) U^\dag_{x,4} \delta_{x,y+\hat4} \right] - \kappa_Q \left[ c_B \sum_{\mu,\nu} F_{\mu\nu}(x) \sigma_{\mu\nu} \right. \\
					& \left. + c_E \sum_{\mu} F_{\mu 4}(x) \sigma_{\mu 4} \right] \delta_{xy},
				\end{split}
			\end{align}
			 with the free parameters $r_s$, $\nu$, $c_B$ and $c_E$ to be tuned in order to remove the discretization errors appropriately. We adopt the perturbative estimates $r_s = 1.1881607$, $c_B = 1.9849139$ and $c_E = 1.7819512$~\cite{Aoki:2004271} and the non-perturbatively tuned $\nu = 1.1450511$ value~\cite{Namekawa:2013vu}. We re-tune the charm-quark hopping parameter to $\kappa_Q = 0.10954007$ non-perturbatively so as to reproduce the relativistic dispersion relation for the 1S spin-averaged charmonium state. With these parameters, masses of the $\eta_c$ and the $J/\psi$ are $m_{\eta_c} = 2.984(2) \text{ GeV}/c^2$, $m_{J/\psi} = 3.099(4) \text{ GeV}/c^2$. The hyperfine splitting is estimated as $\Delta E_{(V-PS)} = 116(4) \text{ MeV}/c^2$, in agreement with its experimental value. Further details of our charm quark tuning can be found in Ref.~\cite{PhysRevD.98.114505}. 

		\subsection{Baryon operators} 
			The baryon operators that we employ are tabulated in \Cref{tab:intop1} in a shorthand notation while the explicit forms of the operators can be found in \Cref{tab:intop2}. Note that we do not distinguish between $u$ and $d$ quarks since they are degenerate in our lattice setup.
			\begin{table}[!h]
				\caption{ \label{tab:intop1} Types of the interpolating operators used for the charmed baryons. Their quark contents are shown in the third columns. }
				\setlength{\extrarowheight}{2pt}
				\centering
				\begin{tabular}{ccc|ccc}
				\hline\hline
				    \multicolumn{3}{c|}{spin-$1/2$} & \multicolumn{3}{c}{spin-$3/2$}\\
				    \hline
					Baryon 	& Operator 	& $(q_1, q_2, q_3)$ & Baryon & Operator & $(q_1, q_2, q_3)$\\
					\hline	
					$\Lambda_c$	& $\Lambda$ - like & $(u,d,c)$ & $\Sigma_c^\ast$ & $\Delta^+$ - like & $(u/d, u/d, c)$ \\
					$\Sigma_c$ & $N$ - like & $(u/d, c, u/d)$ & $\Xi_c^\ast$ & $\Delta^+$ - like & $(u/d, s, c)$ \\
					$\Xi_c$	& $N$ - like & $(u/d, s, c)$ & $\Omega_c^\ast$ & $\Delta^+$ - like & $(s, s, c)$	\\
					$\Xi_c$ & $\Lambda$ - like & $(s, u/d, c)$  \\
					$\Xi_c^\prime$ & $\Xi_c^\prime$ & $(u/d, c, s)$	& $\Xi_{cc}^\ast$ & $\Delta^+$ - like & $(u/d, c, c)$ \\
					$\Omega_c$ & $N$ - like & $(s, c, s)$ & $\Omega_{cc}^\ast$ & $\Delta^+$ - like & $(s, c, c)$ \\
					$\Xi_{cc}$ & $N$ - like & $(c, u/d, c)$	 \\
					$\Omega_{cc}$ & $N$ - like & $(c, s, c)$ & $\Omega_{ccc}$ & $\Delta^+$ - like & $(c, c, c)$ \\
				\hline\hline
				\end{tabular}
			\end{table}

			\begin{table*}[!htb]
			\caption{ \label{tab:intop2} Interpolating operators with generic Dirac structures for spin-$1/2$ and spin-$3/2$ baryons. $C=\gamma_2 \gamma_4$ is the charge conjugation operator. $[\Gamma_1, \Gamma_2]$ choices and the quark contents are given in the text and in \Cref{tab:intop1}.}
			\centering
			\setlength{\extrarowheight}{3pt}
			\resizebox{\textwidth}{!}{
			\begin{tabular}{ccc}
			\hline\hline
				Spin & Baryon & Operator \\
				\hline
				\multirow{4}{*}{1/2} & $N$ - like & $ \varepsilon_{abc} \left[ q_1^{Ta}(x) C \Gamma_1 q_2^b(x) \right] \Gamma_2 q_3^c(x)$ \\
				
				& $\Lambda$ - like & $ \frac{1}{\sqrt{6}} \varepsilon_{abc} \left( 2 \left[ q_1^{Ta}(x) C \Gamma_1 q_2^b(x) \right] \Gamma_2 q_3^c(x) + \left[ q_1^{Ta}(x) C \Gamma_1 q_3^b(x) \right] \Gamma_2 q_2^c(x) - \left[ q_2^{Ta}(x) C \Gamma_1 q_3^b(x) \right] \Gamma_2 q_1^c(x)\right) $ \\

				& $\Xi_c^\prime$ & $ \frac{1}{\sqrt{2}} \varepsilon_{abc} \left( \left[ q_1^{Ta}(x) C \Gamma_1 q_2^b(x) \right] \Gamma_2 q_3^c(x) + \left[ q_3^{Ta}(x) C \Gamma_1 q_2^b(x) \right] \Gamma_2 q_1^c(x) \right) $ \\
				
				3/2 & $\Delta^+$ - like & $ \frac{1}{\sqrt{3}} \varepsilon_{abc} \left( 2 \left[ q_1^{Ta}(x) C \gamma_\mu q_2^b(x) \right] q_3^c(x) + \left[ q_1^{Ta}(x) C \gamma_\mu q_3^b(x) \right] q_2^c(x) \right) $ \\
			\hline\hline
			\end{tabular}}
			\end{table*}           

			For the spin-$1/2$ baryon, we form three individual operators by using the Dirac structures, $[\Gamma_1, \Gamma_2] = [\gamma_5,1]$, $[1,\gamma_5]$, and $[\gamma_5 \gamma_4,1]$ (see \Cref{tab:intop2}). An explicit example for the $N$-like operator is
			\begin{align}
				\chi_1(x) &= \varepsilon_{abc} \left[ q_1^{Ta}(x) C \gamma_5 q_2^b(x) \right] q_3^c(x), \\
				\chi_2(x) &= \varepsilon_{abc} \left[ q_1^{Ta}(x) C q_2^b(x) \right] \gamma_5 q_3^c(x), \\
				\chi_4(x) &= \varepsilon_{abc} \left[ q_1^{Ta}(x) C \gamma_5 \gamma_4 q_2^b(x) \right] q_3^c(x).
			\end{align}
			The $\chi_4$-type operator with the Dirac structure $[\Gamma_1, \Gamma_2] = [\gamma_5 \gamma_4,1]$ corresponds to the time component of an operator with $[\Gamma_1, \Gamma_2] = [\gamma_5 \gamma_\mu,\gamma_5]$, which couples to both spin-$1/2$ and spin-$3/2$ particles. It has been shown that projecting out the spin-$1/2$ component of such an operator results in two terms: a linear combination of the $\chi_1$ and the $\chi_2$, and a term containing the $\chi_4$ operator~\cite{PhysRevD.76.054510}. Furthermore, the $\chi_4$-type operator is distinct from the $\chi_1$ and the $\chi_2$ from a chiral transformation perspective~\cite{PhysRevD.69.094513}, making it a viable choice for the basis set of the spin-$1/2$ operators.   
			
			We limit ourselves to only one Dirac structure for the spin-$3/2$ baryons, which is $[\Gamma_1, \Gamma_2] = [\gamma_\mu ,1]$. Note that if one uses $N$~-~like operators for spin-$3/2$ baryons, there would be a mixing coming from the corresponding spin-$1/2$ states. In that case, it would be necessary to project the individual interpolating operators to definite spin-$3/2$ states in order to remove such contaminations. On the other hand, $\Delta$~-~like operators that we use already have a good overlap to spin-$3/2$ states with negligible spin-$1/2$ components. Mixing between the spin-$3/2$ and spin-$1/2$ states has been studied in detail in Ref.~\cite{Alexandrou:2014sha} for the strange and charmed baryons where it has been shown that a spin-$3/2$ projection is indeed not necessary for $\Delta$~-~like operators.
			
			Among the operators discussed in this section, the ones coupling to the $\Xi_c$ and $\Xi_c^\prime$ states deserve special attention. 
			The $\Xi_c$ ($\Xi_c^\prime$), which belongs to an $SU(3)$ anti-triplet (sextet) is anti-symmetric (symmetric) with respect to the exchange 
			of $s$ and $u/d$ quarks, which should hold for the respective operators. For $\Xi_c$, this can be achieved by both $N$-like and $\Lambda$-like 
			operators, which will both be used in this work. 
			Note that our $N$-like $\Xi_c$ operator was referred to as ``HQET" in Ref.~\cite{PhysRevD.92.034504}. 
			For $\Xi_c^\prime$, we employ a different operator combination with the correct symmetry properties as shown in \Cref{tab:intop2}. 
			While $\Xi_c$ and $\Xi_c^\prime$ states decouple in the $SU(3)$ limit, they can in principle mix in our setup due to the breaking of the $SU(3)$ symmetry. This mixing can be studied by computing cross-correlators of $\Xi_c$ and $\Xi_c^\prime$ operators. The results of such an analysis will be discussed in Section~\ref{sec:rd}.

		\subsection{Simulation details} 
			Quark fields of the interpolating operators are Gaussian smeared in a gauge-invariant manner at the source, $(x,y,z,t)=(16a,16a,16a,16a)$, for all the baryons with three different sets of smearing parameters, corresponding to an rms radius of $\sim 0.2$, $0.4$ and $0.7$ fms for the quark wave-functions. Sink operators are smeared in the same manner. However, we find that the signal deteriorates rapidly with increasing sink operator smearing. For this reason, we analyze the spin-$1/2$ baryons with smeared-source -- point-sink correlation functions with a fixed source smearing for all the quark fields. Correlation functions depend mildly to the smearing of the singly represented quarks and the plateau regions become independent of the smearing after a certain number of iterations. Therefore, we apply the smearing to the quark fields depending on their flavor and quantity. We treat the $u$-, $d$- and the $s$-quarks on an equal footing and consider them as \emph{light} quarks in comparison to the charm quark. When the interpolating operator is formed by two light quarks and a charm quark, we fix the smearing of the charm quark to $0.7$ fms, which is the widest of the smearings that we have, to decouple its effects and perform the variational analyses over the smearings of the remaining light quarks. Smearing parameters of the individual light quarks are set to be equal. This is true for all the baryon fields with the exceptions of $\Omega^{(\ast)}_{cc}$, in which case the smearing of the strange quark is fixed to $0.7$ fms and the smearings of the charm quarks are varied, and $\Omega_{ccc}$, for which the treatment is the same as light quarks. For the spin-$3/2$ baryons, we use smeared-source -- smeared-sink correlators to form an operator basis from an operator with fixed Dirac structure. A discussion on the operator basis is given in \Cref{sec:opDep}. Parity is selected by applying the parity projection operator, $P^{\pm}$, to the individual correlation functions.

			We bin our data with a bin size of $15$ measurements to account for the autocorrelations on this ensemble and estimate the statistical errors via a single elimination jackknife analysis. We performed our computations using a modified version of the Chroma software system~\cite{Edwards:2004sx} on CPU clusters along with the QUDA library~\cite{Babich:2011np,Clark:2009wm} for the valence $u-/d-$ and $s$-quark propagator inversions on GPUs. The charm quark inversions are done on CPUs. 

		\section{Results And Discussion}\label{sec:rd}
			\subsection{Variational analysis}
\begin{figure*}[htb]
					\centering
					\includegraphics[width=\textwidth]{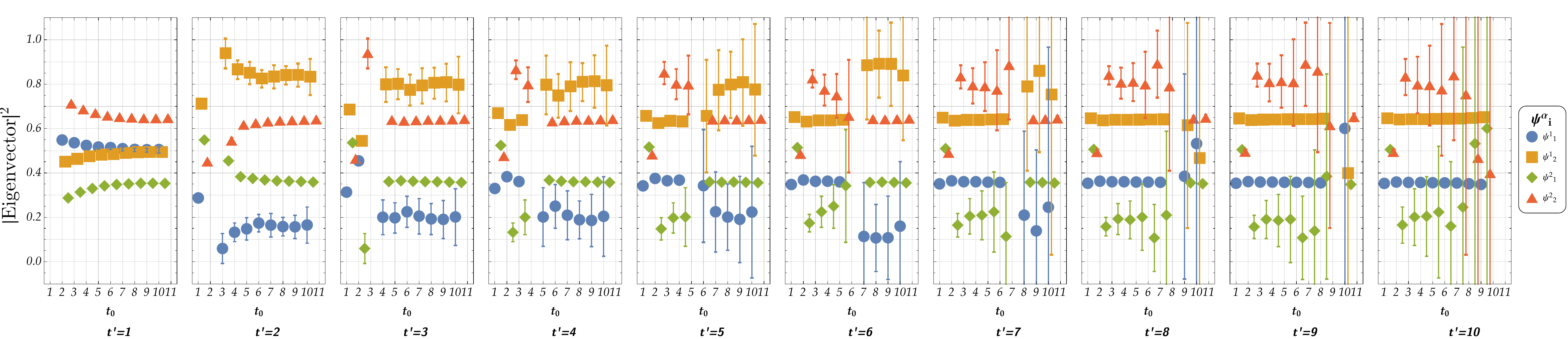} \\
					\includegraphics[width=\textwidth]{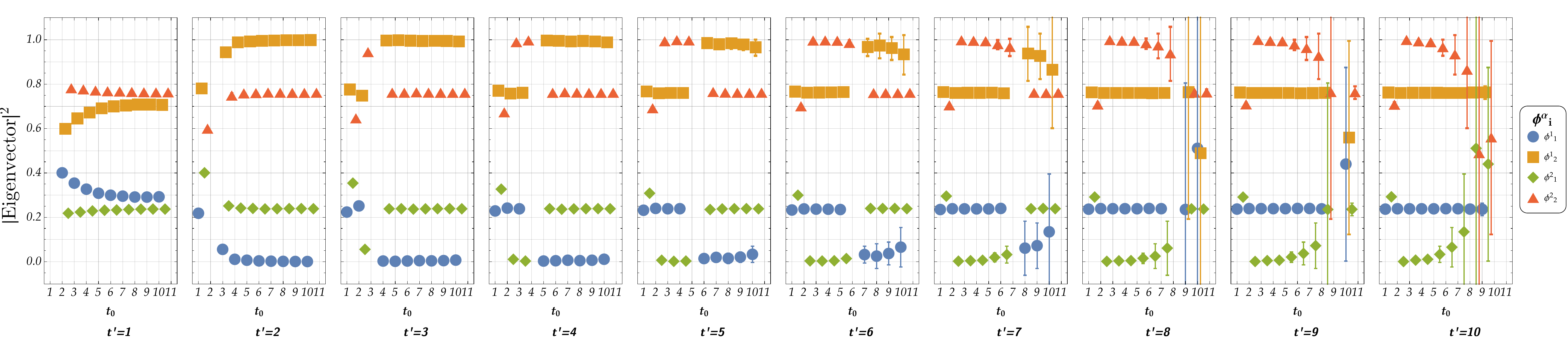}
					\caption{ \label{fig:cchk} Consistency check of the variational parameters for positive parity spin-$1/2$ $\Xi_c$ with $\sim0.7$ fm quark field smearing. Plots show the left and right eigenvectors, $\psi^\alpha$ and $\phi^\alpha$ for varying reference time, $t_0$, and the time-slice of the eigenvector, $t^\prime$. Association of the operators to the states flip when $t_0 > t^\prime$.}
				\end{figure*}

\begin{figure*}[htb]
					\centering
					\includegraphics[width=.49\textwidth]{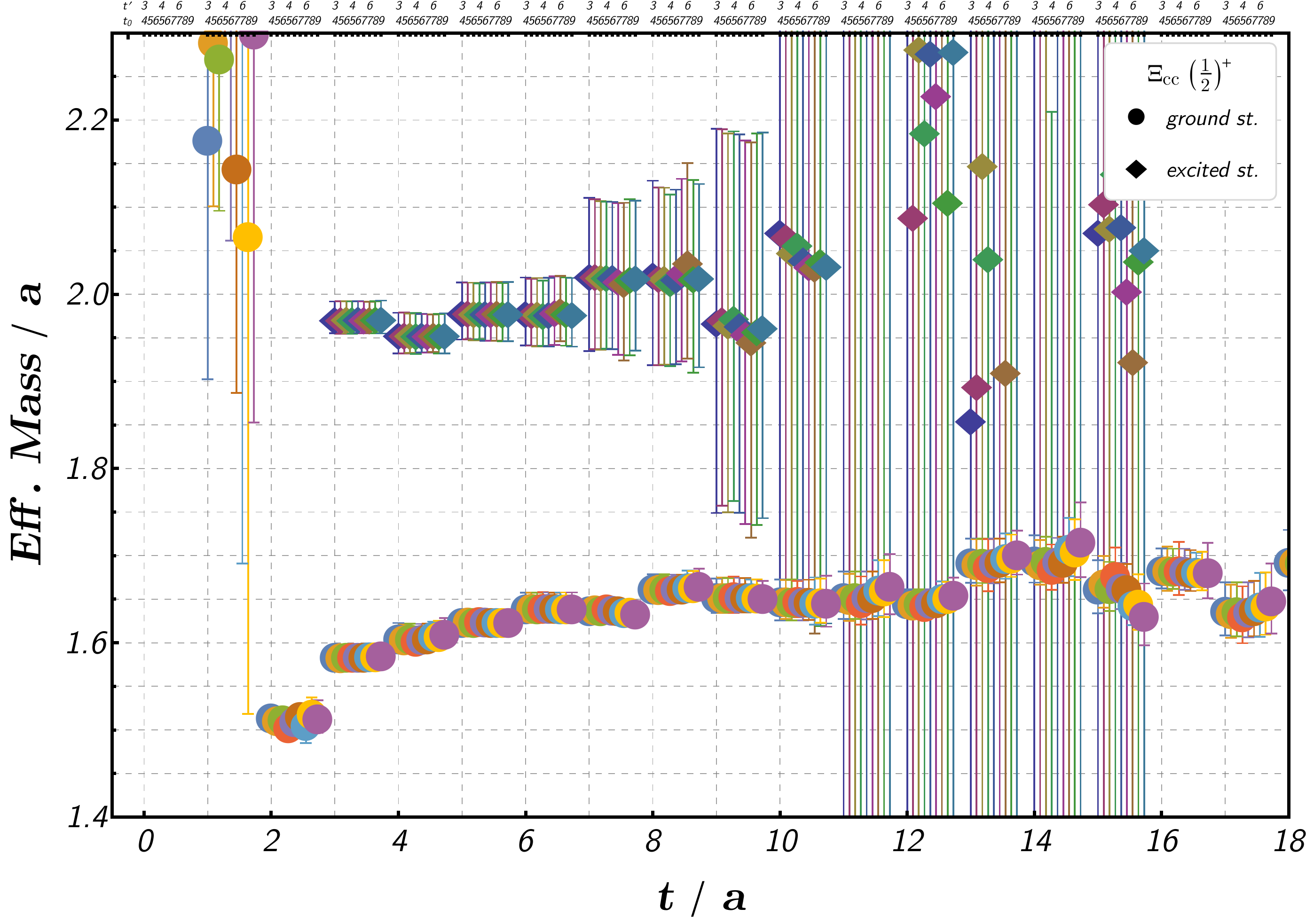}
					\includegraphics[width=.49\textwidth]{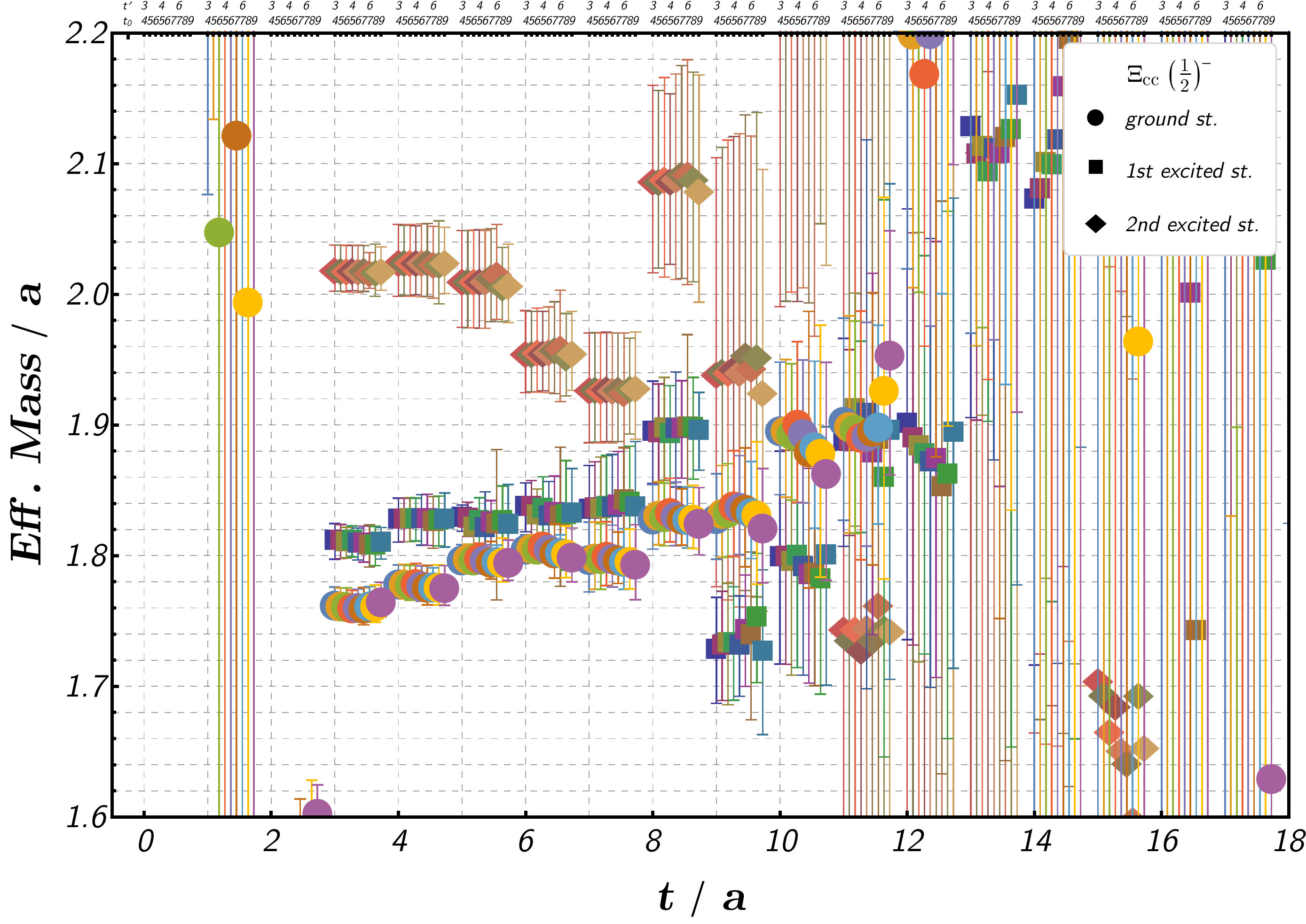}
					\caption{ \label{fig:effmass_cchk} Effective mass plots for the diagonalized correlation matrices (\Cref{eq:dc}) constructed from the solutions of the generalized eigenvalue problem (\Cref{eq:gev}) with a range of variational parameters $ 3a \le t^\prime \le 6a$ and $ t^\prime < t_0 \le t^\prime + 3a$ for each $t^\prime$. An illustrative case for the positive (left) and negative (right) parity spin-$1/2$ $\Xi_{cc}$ states is shown. 
					}
				\end{figure*}

				To obtain the individual states from a set of operators, one solves the generalized eigenvalue problem on each time slice, $t$, against a reference time-slice, $t_0$, as discussed in Section\,\ref{sec:lf}. To ensure the consistency of this step, it is necessary to check that the solutions are stable with respect to $t_0$, since it can be chosen freely. Another concern is associating the eigenvalues with the states. Eigenvalues are sorted in increasing order on each time slice. However due to the faster deterioration of the higher states' signal, their eigenvalues fluctuate heavily as time evolves and can sometimes be smaller than the eigenvalue associated with the lower state. This situation might misguide the analysis if not addressed properly. In order to make sure that the eigenvalues are associated with the correct states, we fix the time-slice of the eigenvectors, $t^\prime$, that is used to diagonalize the correlation function matrix, to a specific value. This procedure, however, introduces an extra parameter dependence to the analysis. We check this dependence for each channel for a range of $t^\prime$ values. The dependencies on $t_0$ and $t^\prime$ can be tracked by investigating the respective eigenvectors, whose components should be stable when changing both fictitious time parameters. We illustrate such a consistency check in \Cref{fig:cchk,fig:effmass_cchk}. We perform this check for each channel and select a $(t^\prime,t_0)$ combination, where $t^\prime \ge 2 a$ and $t_0 > t^\prime$, that optimizes the signal quality.

				\subsubsection{Operator dependence}\label{sec:opDep}
\begin{figure*}[htb]
						\centering
						\subfloat[][$\Xi_{cc} \left( \frac{1}{2}^+ \right)$]{
							\label{fig:opComp_xicc12p} 
							\includegraphics[width=.325\textwidth]{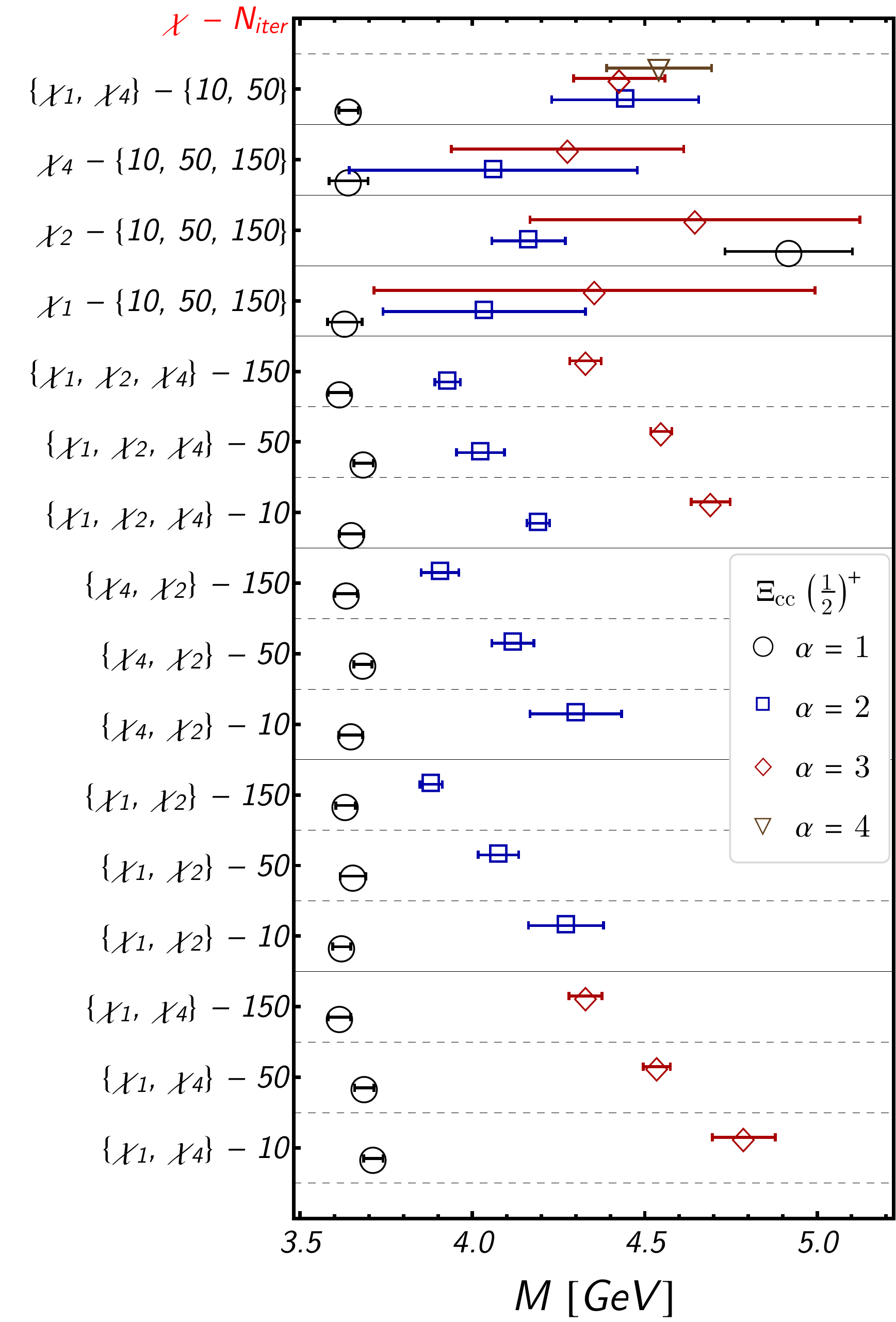}
						}
						\subfloat[][$\Omega_{c} \left( \frac{1}{2}^+ \right)$]{
							\label{fig:opComp_omegac12p}
							\includegraphics[width=.325\textwidth]{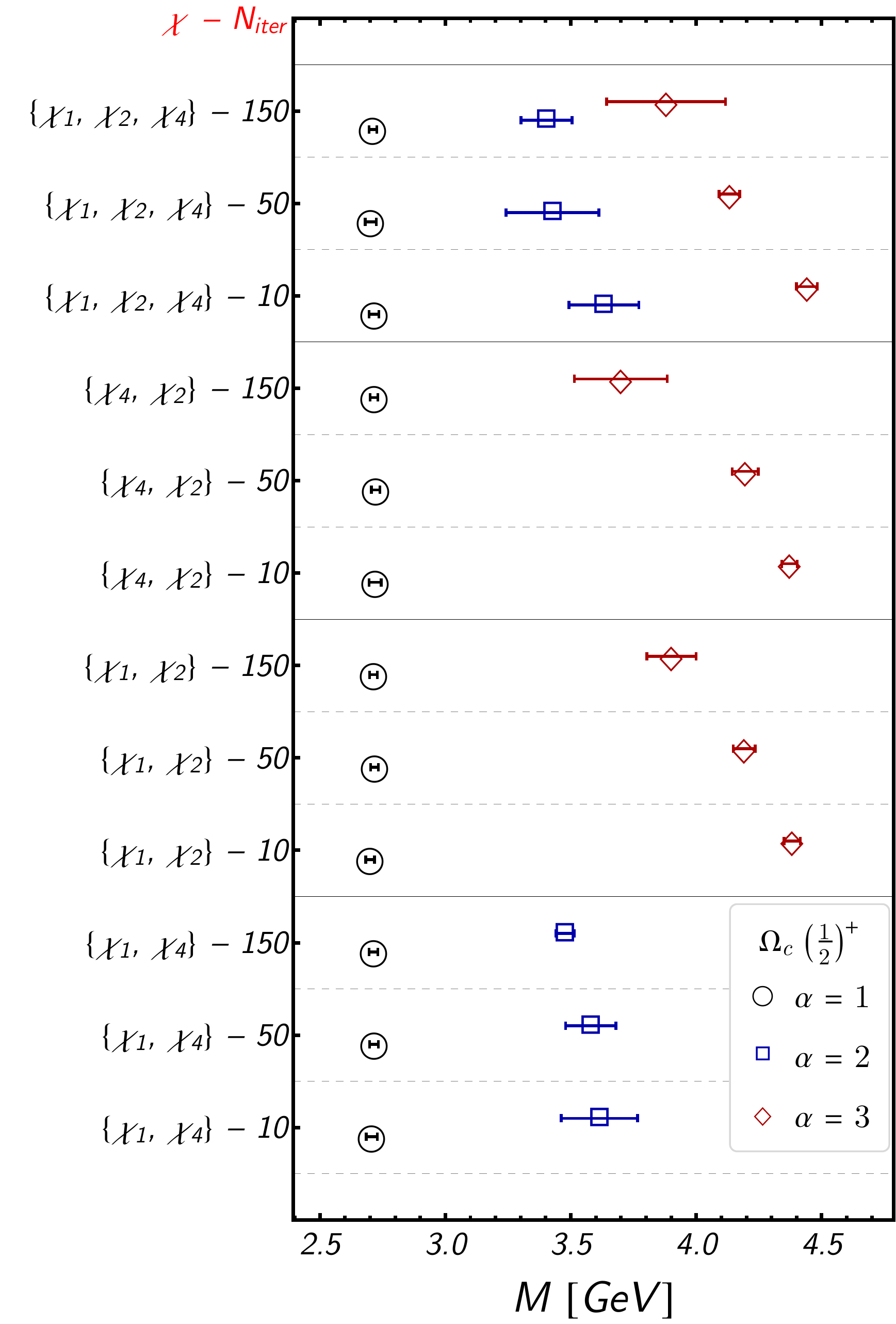}
						}
						\subfloat[][$\Xi_{c} \left( \frac{1}{2}^+ \right)$]{
							\label{fig:opComp_xic12p}
							\includegraphics[width=.325\textwidth]{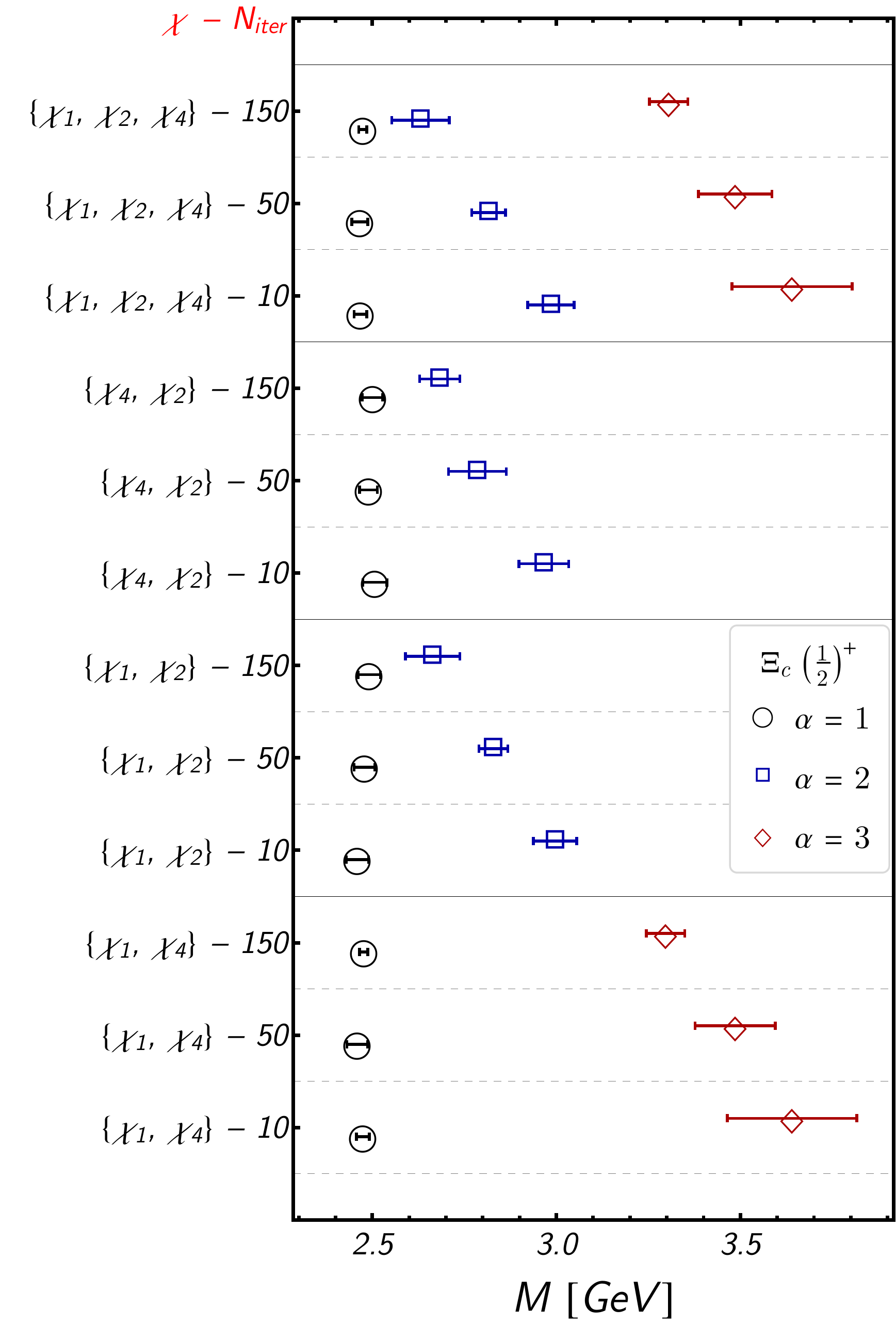}
						} \\ \vspace{2mm}
						\subfloat[][$\Xi_{cc} \left( \frac{1}{2}^- \right)$]{
							\label{fig:opComp_xicc12n}
							\includegraphics[width=.325\textwidth]{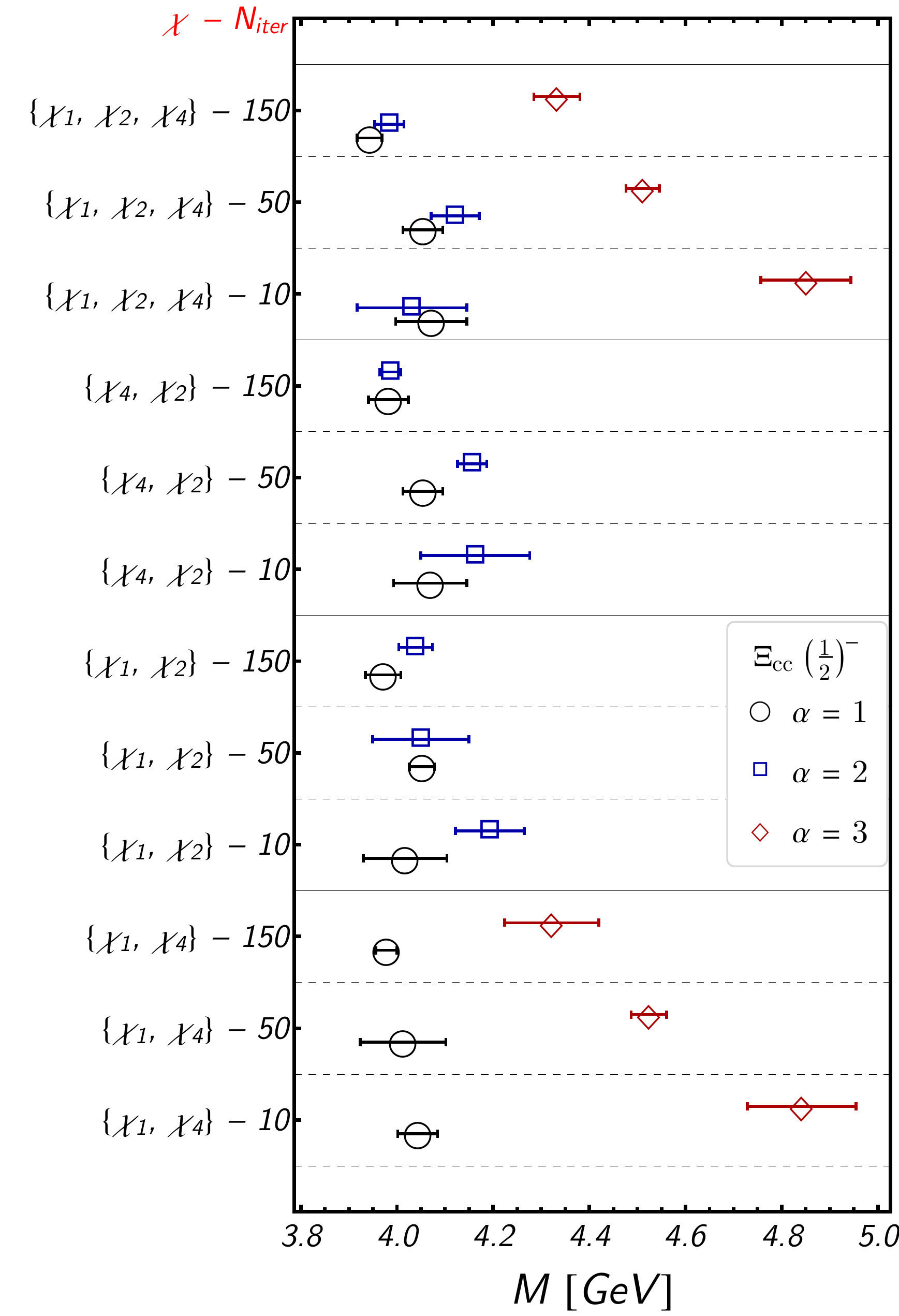}
						}
						\subfloat[][$\Omega_{c} \left( \frac{1}{2}^- \right)$]{
							\label{fig:opComp_omegac12n}
							\includegraphics[width=.325\textwidth]{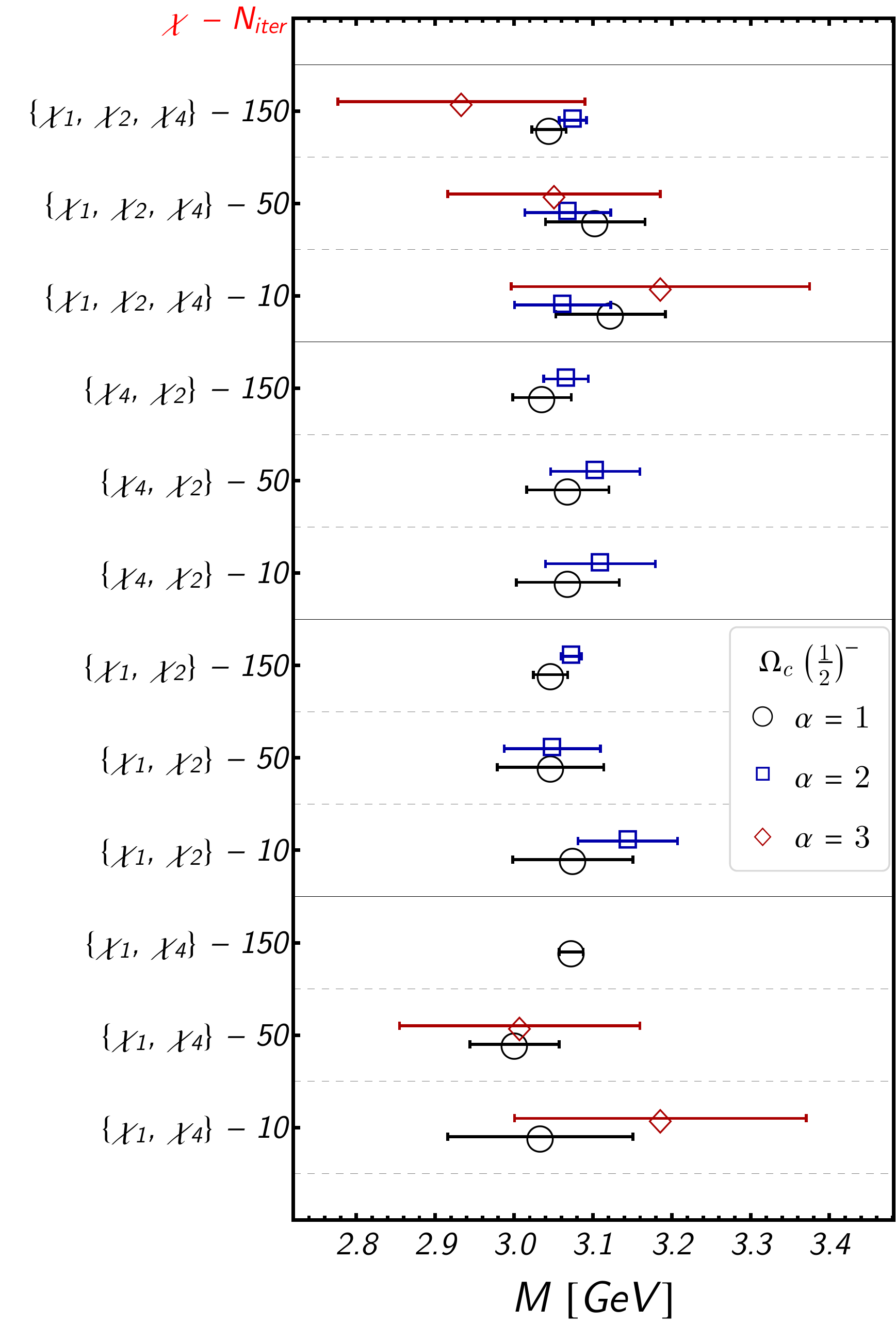}
						}
						\subfloat[][$\Xi_{c} \left( \frac{1}{2}^- \right)$]{
							\label{fig:opComp_xic12n}
							\includegraphics[width=.325\textwidth]{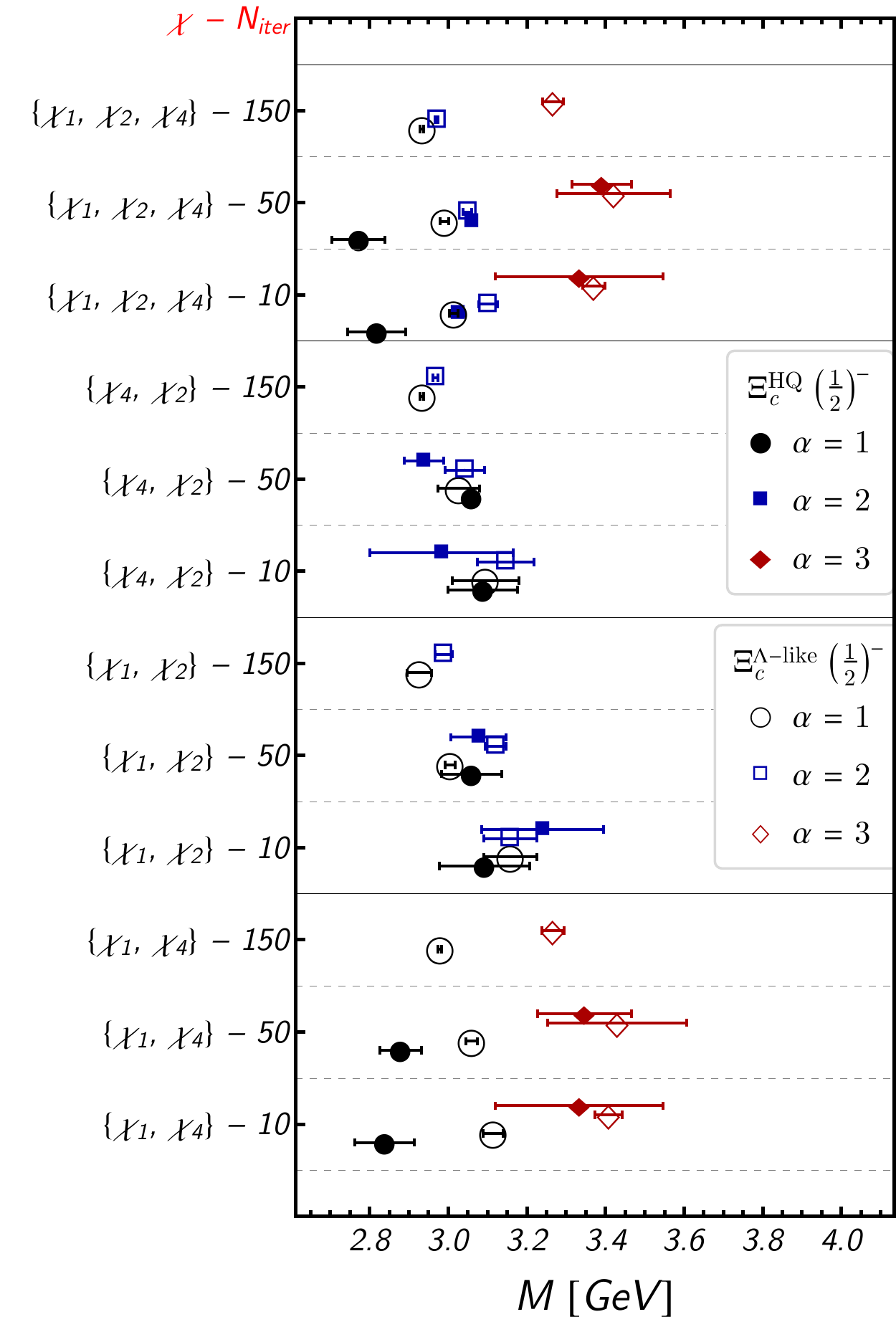}
						}
						\caption{ \label{fig:opComp} Operator dependence of the extracted states for representative baryon channels. Vertical axes label the operator basis with respect to the Dirac structure, $\chi$, and the smearing steps of the quark fields, $N_\text{iter}$, where the iterations correspond to an rms radius of $\sim 0.2$, $0.4$ and $0.7$ fms for increasing $N_\text{iter}$. Data points in each section, divided by dashed or solid lines, are shifted for clarity. Filled symbols in \protect\subref{fig:opComp_xic12n} correspond to the states extracted via an $N$-like operator basis given in \Cref{tab:intop2}. Note that we only have two smearings for that case. State numbering, $\alpha=1 - 3$, follows the notation of the $3\times3$ solutions even for the $2\times2$ solutions to emphasize the coupling of certain operators to certain states. All energies are extracted via a plateau method, see main text for a discussion. } 
					\end{figure*}

					\paragraph{Operator basis:} Having three operators with differing Dirac structures, it is possible to analyze both the full $3 \times 3$ correlator, but also various combinations of $2 \times 2$ correlators. While the full information for all of them is contained in the $3 \times 3$ case, the $2 \times 2$ correlators can provide valuable and comprehensible information about which state couples to which operator. For this purpose we here investigate the correlators with different operator sets. We find that the variational analyses over two different sets of spin-$1/2$ operators, namely over $\{\chi_1, \chi_4 \}$ and $\{\chi_1, \chi_2 \}$, give two distinct second eigenvalues for the positive parity states. The $\{\chi_4, \chi_2 \}$ set produces similar results to that of $\{\chi_1, \chi_2 \}$. For negative parity, only the $\{\chi_1, \chi_4 \}$ combination yields mostly well-separated second eigenvalues, whereas the second eigenvalues of the $\{\chi_1, \chi_2 \}$ and $\{\chi_4, \chi_2 \}$ bases lie closer to the first eigenvalues. When we extend the operator basis to the $\{\chi_1, \chi_2, \chi_4 \}$ set and solve the corresponding $3 \times 3$ variational system, the $2 \times 2$ results are reproduced. These findings are illustrated in \Cref{fig:opComp} for the positive and negative parity $\Xi_{c}$, $\Omega_c$, and $\Xi_{cc}$ baryons where we show the fit results from a plateau approach.  These representative baryons are chosen such that they correspond to the different operator characteristics, i.e.  $\Lambda$-like, singly-charmed $N$-like, and doubly-charmed $N$-like, respectively.

\begin{figure*}[htb]
						\centering
						\subfloat[][$\Xi_{cc} \left( \frac{1}{2}^+ \right)$]{
							\label{fig:opEvecs_xicc}
							\includegraphics[width=.294\textwidth, trim={0 5mm 17mm 0},clip]{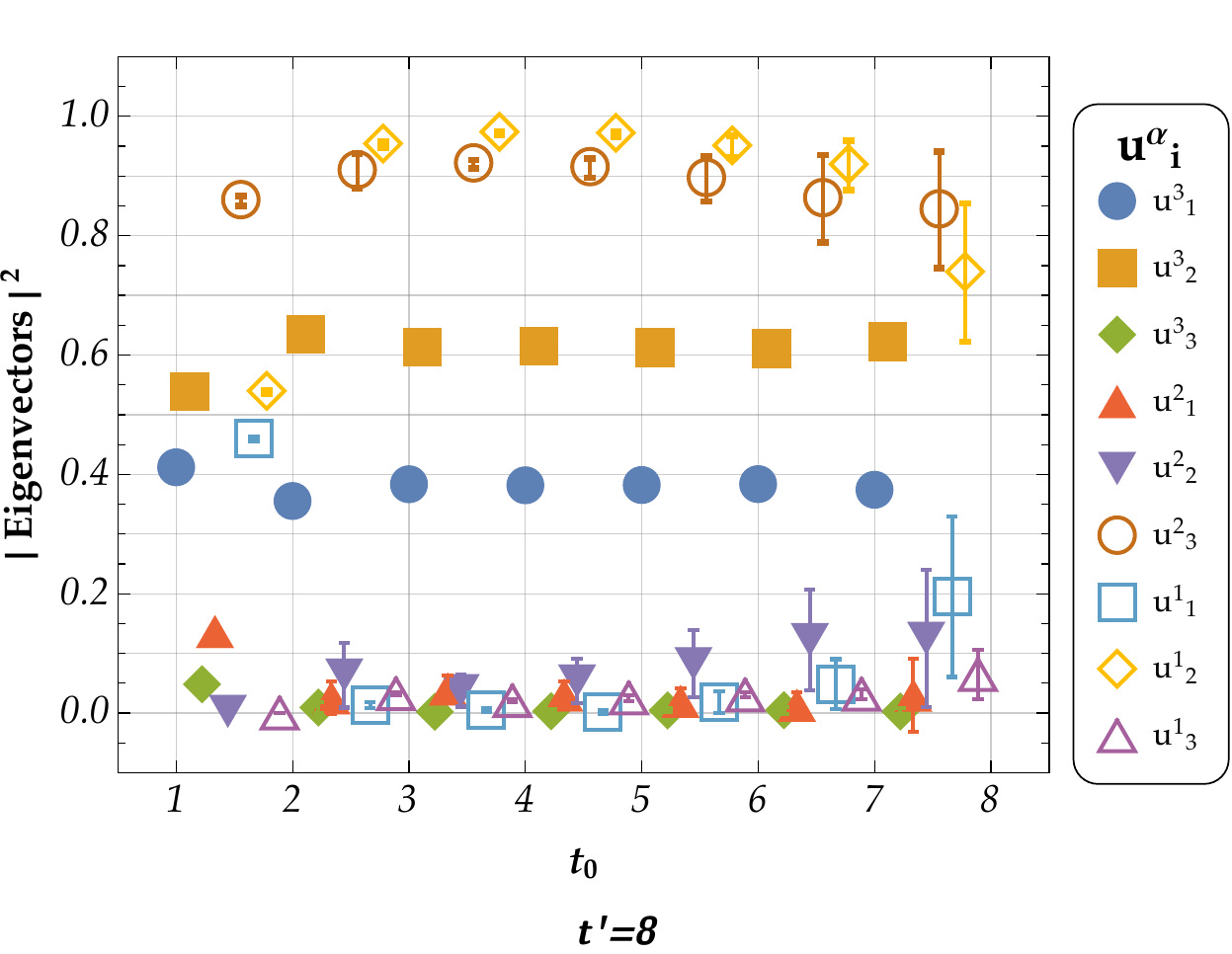}
						} \hfill
						\subfloat[][$\Omega_{c} \left( \frac{1}{2}^+ \right)$]{
							\label{fig:opEvecs_omegac}
							\includegraphics[width=.28\textwidth, trim={5mm 5mm 17mm 0},clip]{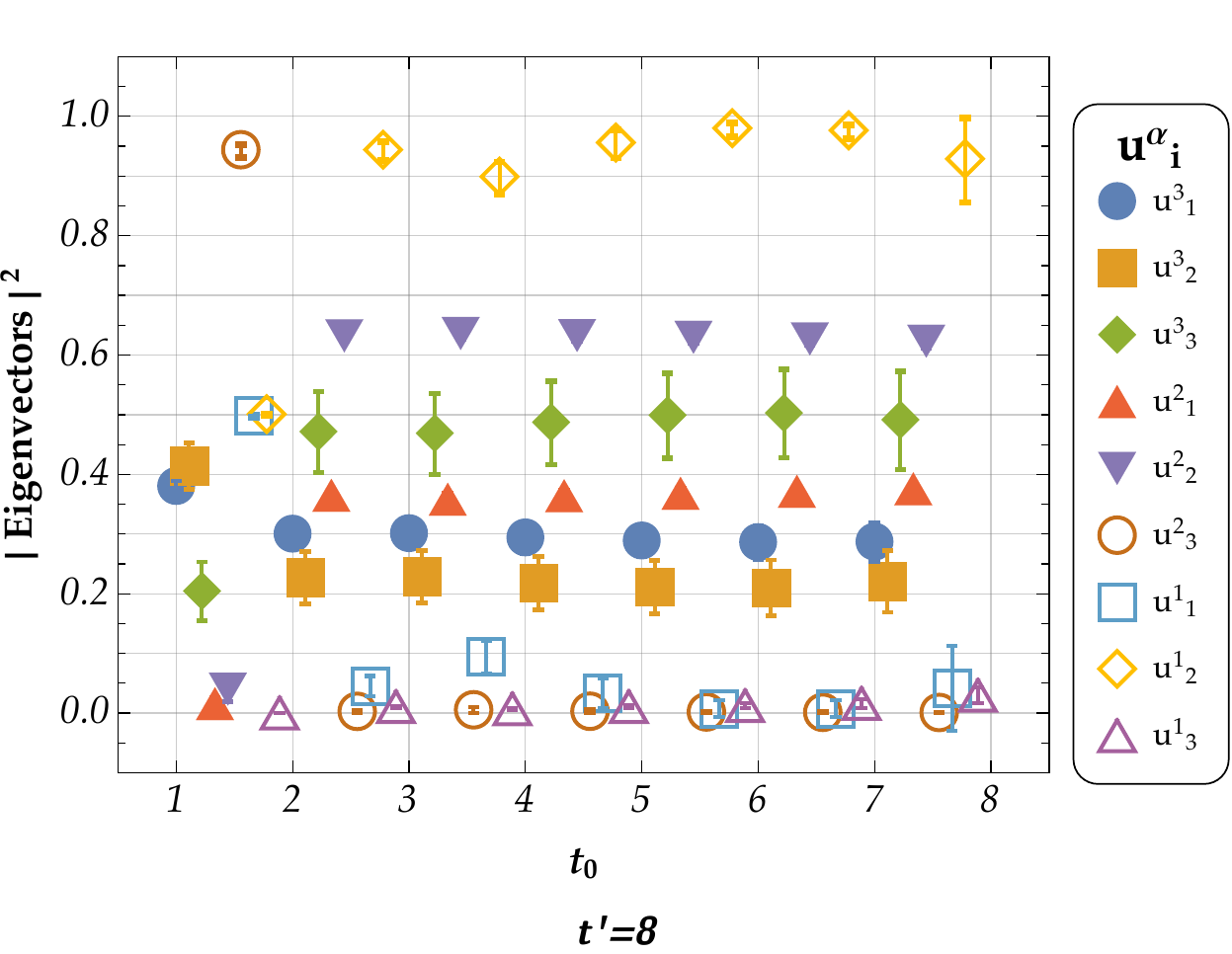} 
						} \hfill
						\subfloat[][$\Xi_{c} \left( \frac{1}{2}^+ \right)$]{
							\label{fig:opEvecs_xic}
							\includegraphics[width=.325\textwidth, trim={5mm 5mm 0 0},clip]{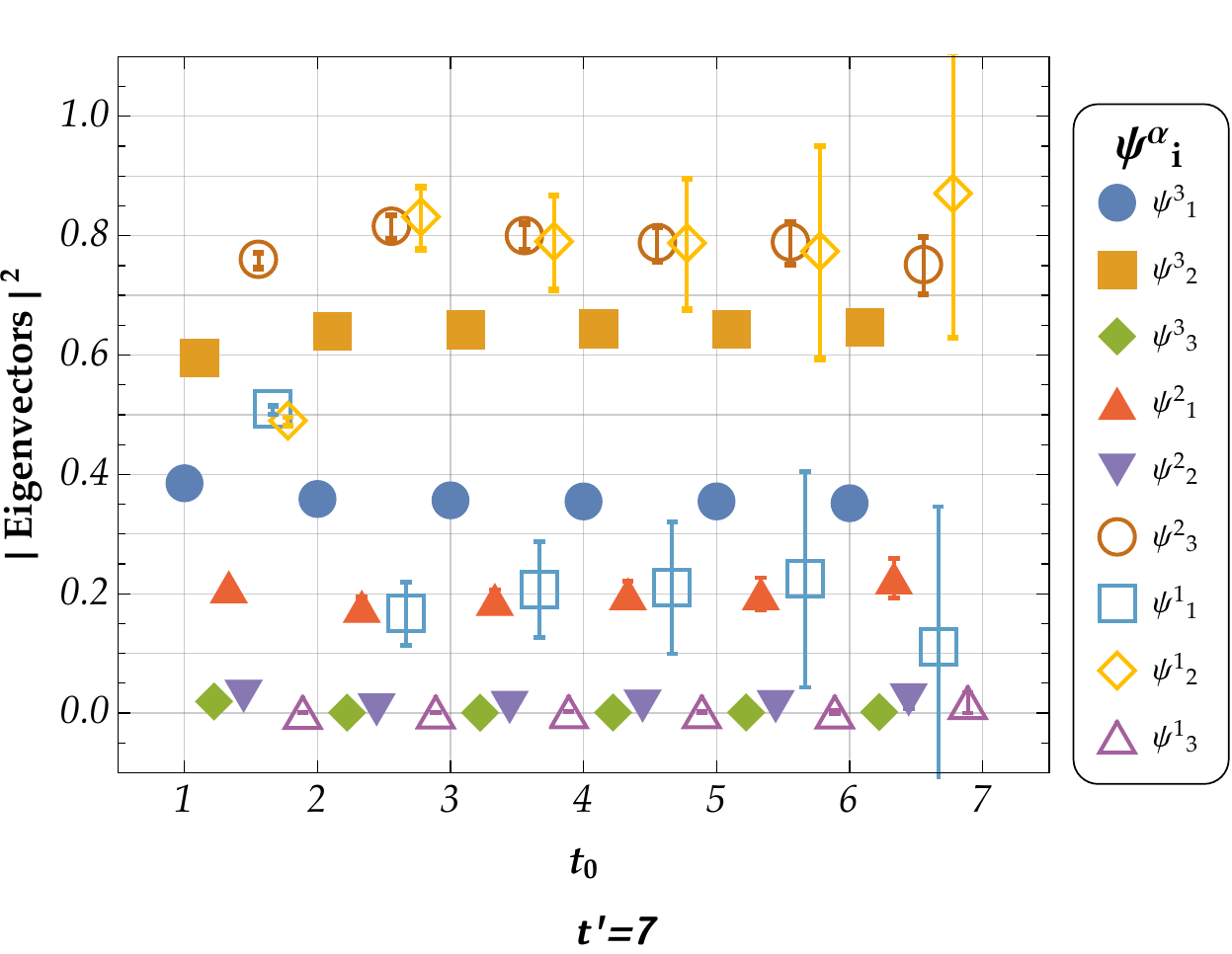}
						}
						\caption{ \label{fig:opEvecs} Eigenvectors from $3 \times 3$ variational solutions for the \protect\subref{fig:opEvecs_xicc} $\Xi_{cc}$, \protect\subref{fig:opEvecs_omegac} $\Omega_c$ and \protect\subref{fig:opEvecs_xic} $\Xi_c$ channels. $\psi^\alpha_i$ is the right eigenvector where $\alpha$ is the state and the index $i$ stands for the individual operator in the operator basis $i=\{\chi_1, \chi_4, \chi_2\}$. For instance $\psi^2_3$ corresponds to the contribution of $i_3=\chi_2$ to the second state. }
					\end{figure*}

					\paragraph{$N$-like operators:} Although we use the same $N$-like operators for the singly-charmed and the doubly-charmed spin-$1/2$ baryons, it is reasonable to expect a different behavior when we solve the variational system, since they belong to different layers of the mixed-flavor $SU(4)$ 20-plet. Such a difference is evident when we compare the solutions from the operator sets $\{\chi_1, \chi_4 \}$, $\{\chi_1, \chi_2 \}$, and $\{\chi_4, \chi_2 \}$. The lower three sections, divided by the solid lines, of the positive parity $\Xi_{cc}$, and $\Omega_{c}$ in \Cref{fig:opComp_xicc12p,fig:opComp_omegac12p} show that different operators couple to different states. The different couplings can be tracked to the eigenvectors of each solution as shown in \Cref{fig:opEvecs}. The $\chi_2$ operators couple only to the second states in the $\Xi_{cc}$ channel while it couples to the third state only for the $\Omega_c$. 

\begin{figure*}[htb]
						\centering	
						\subfloat{\includegraphics[width=.49\textwidth]{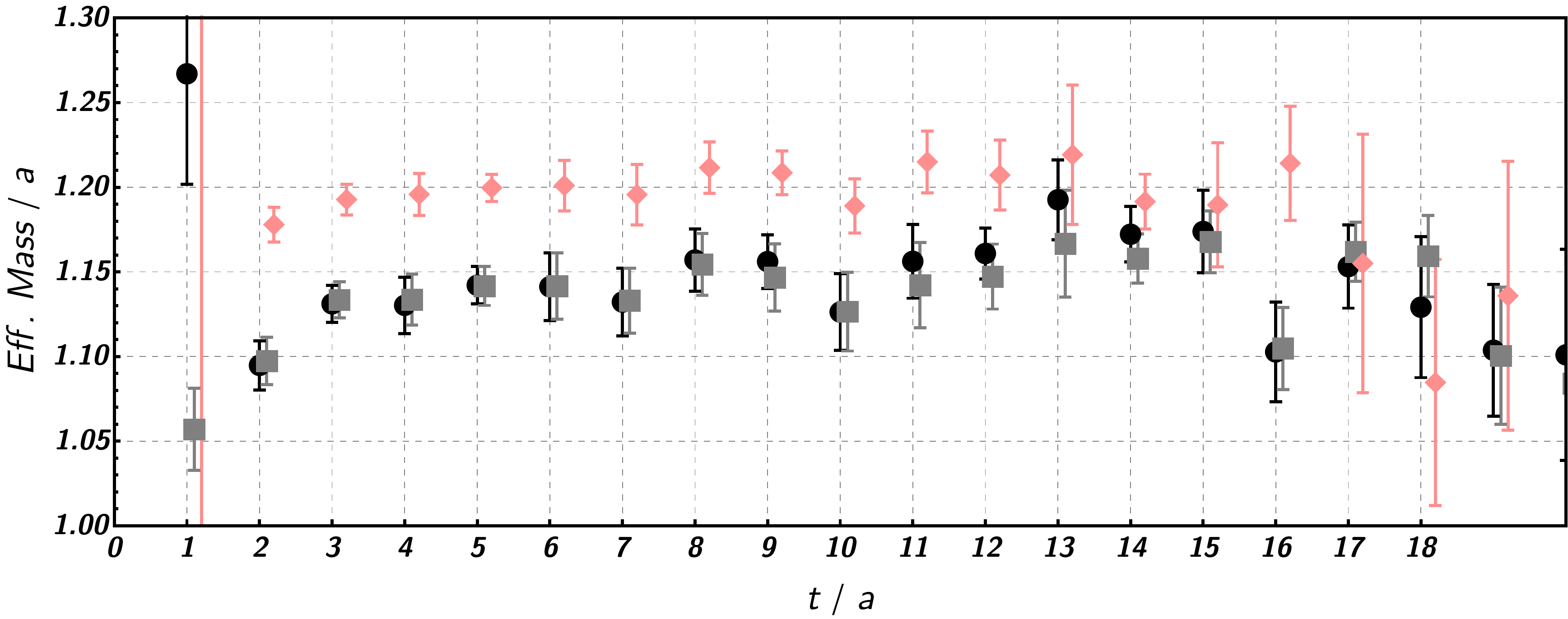} } 
						\hfill
						\subfloat{\includegraphics[width=.49\textwidth]{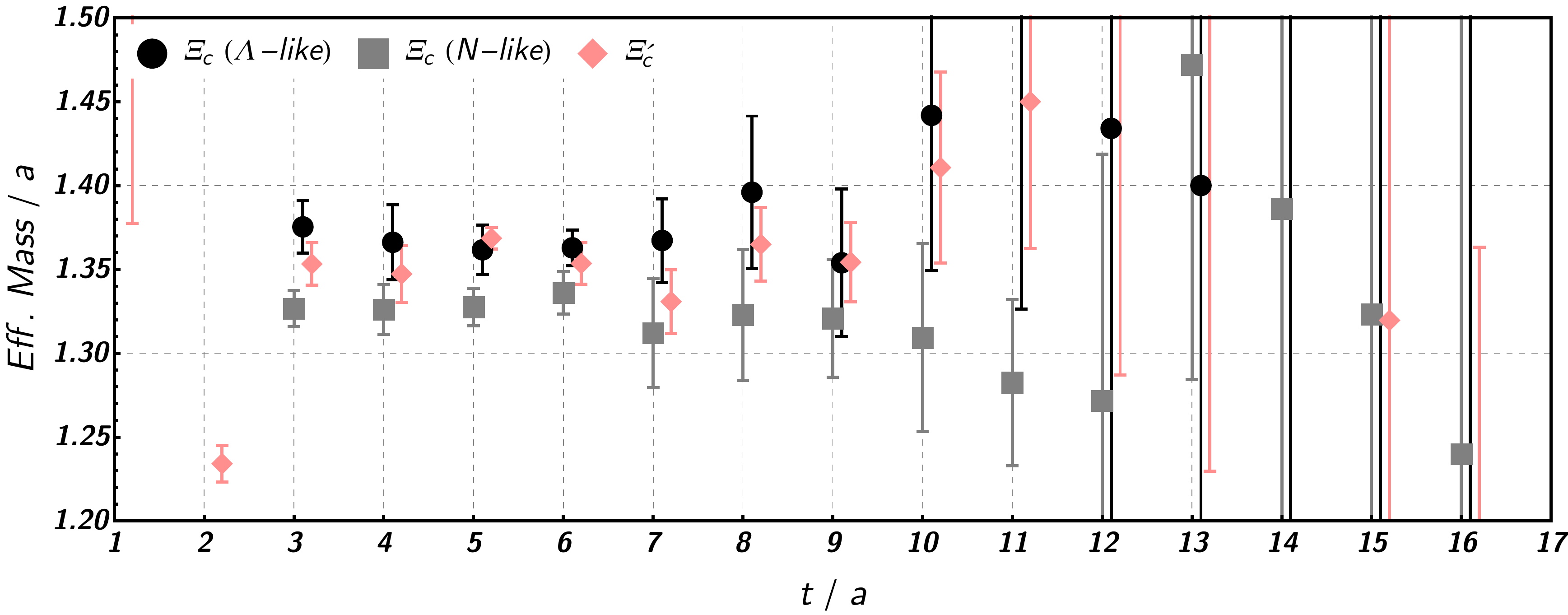} } \\
						\hspace{5mm}
						\subfloat{\includegraphics[width=.23\textwidth]{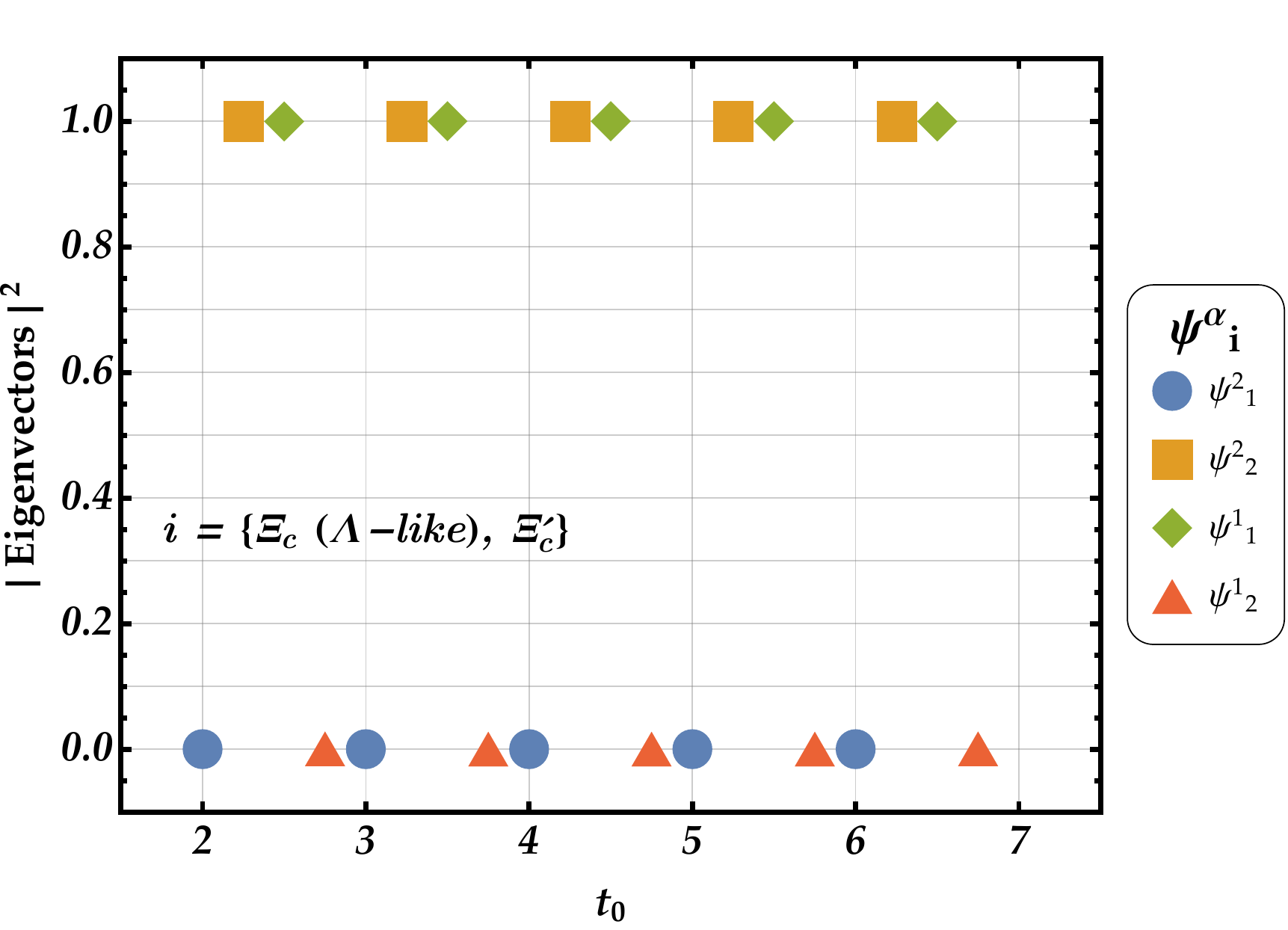} }
						\subfloat{\includegraphics[width=.23\textwidth]{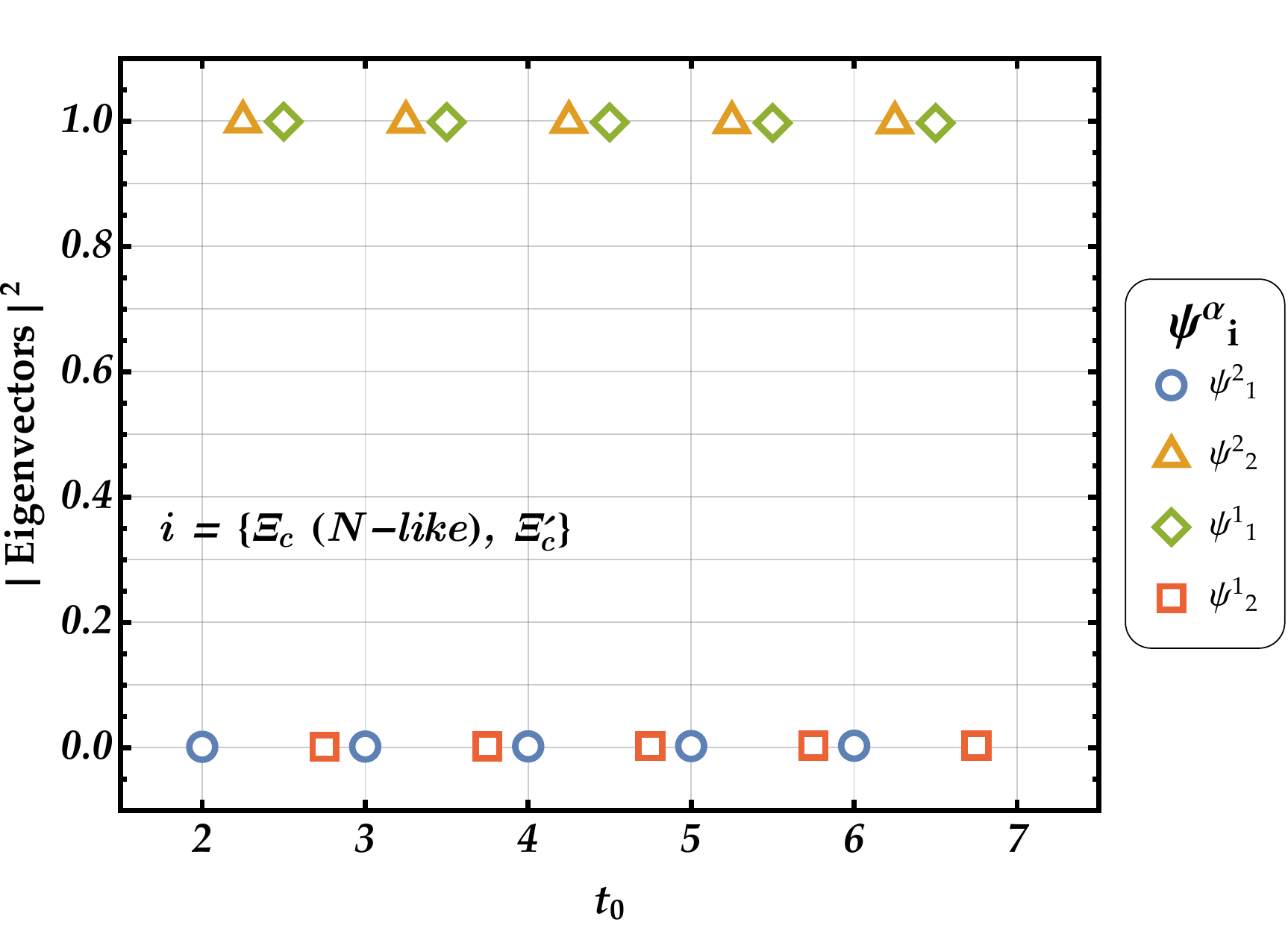} }
						\hspace{3mm}
						\subfloat{\includegraphics[width=.23\textwidth]{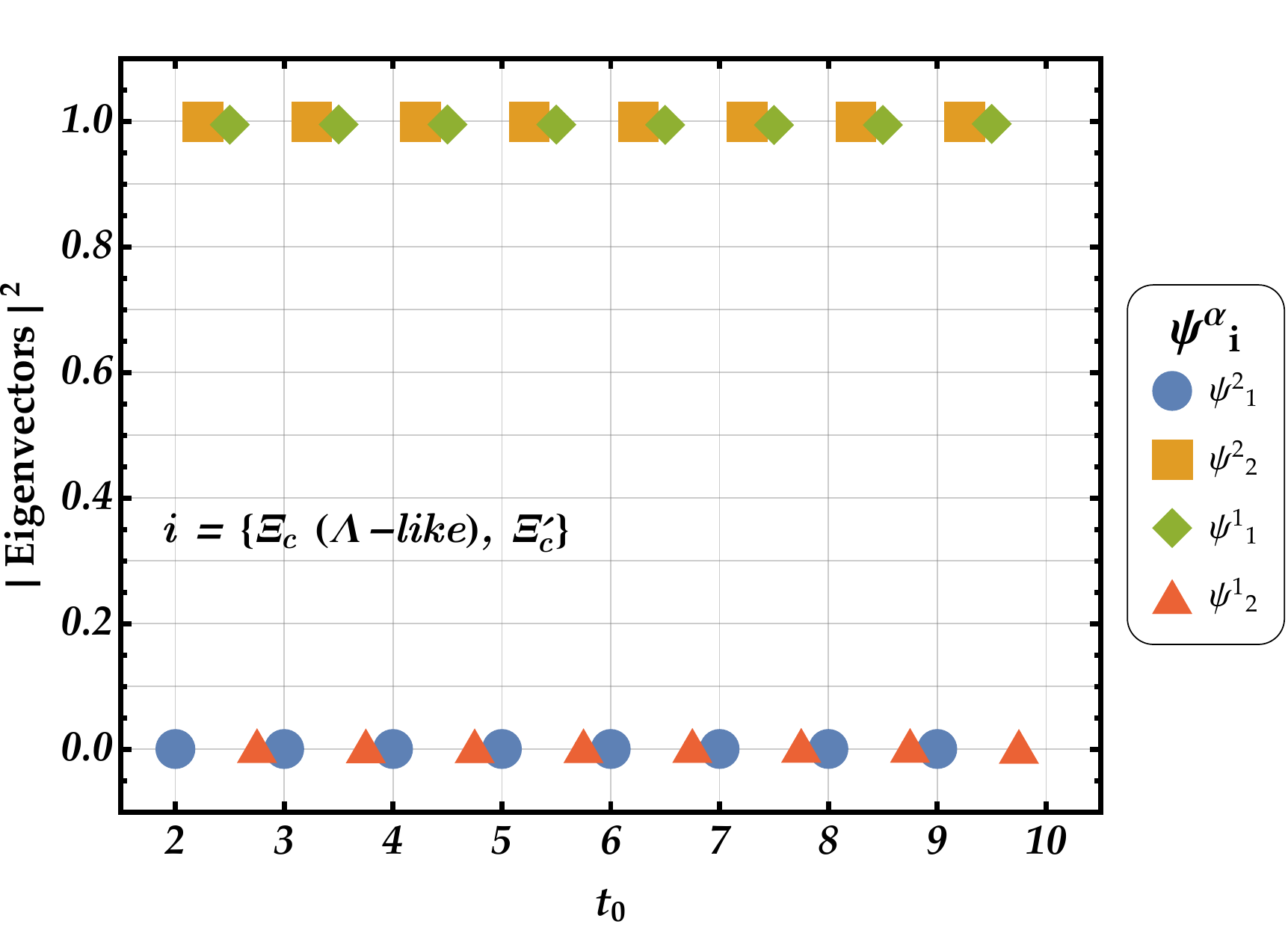} }
						\subfloat{\includegraphics[width=.23\textwidth]{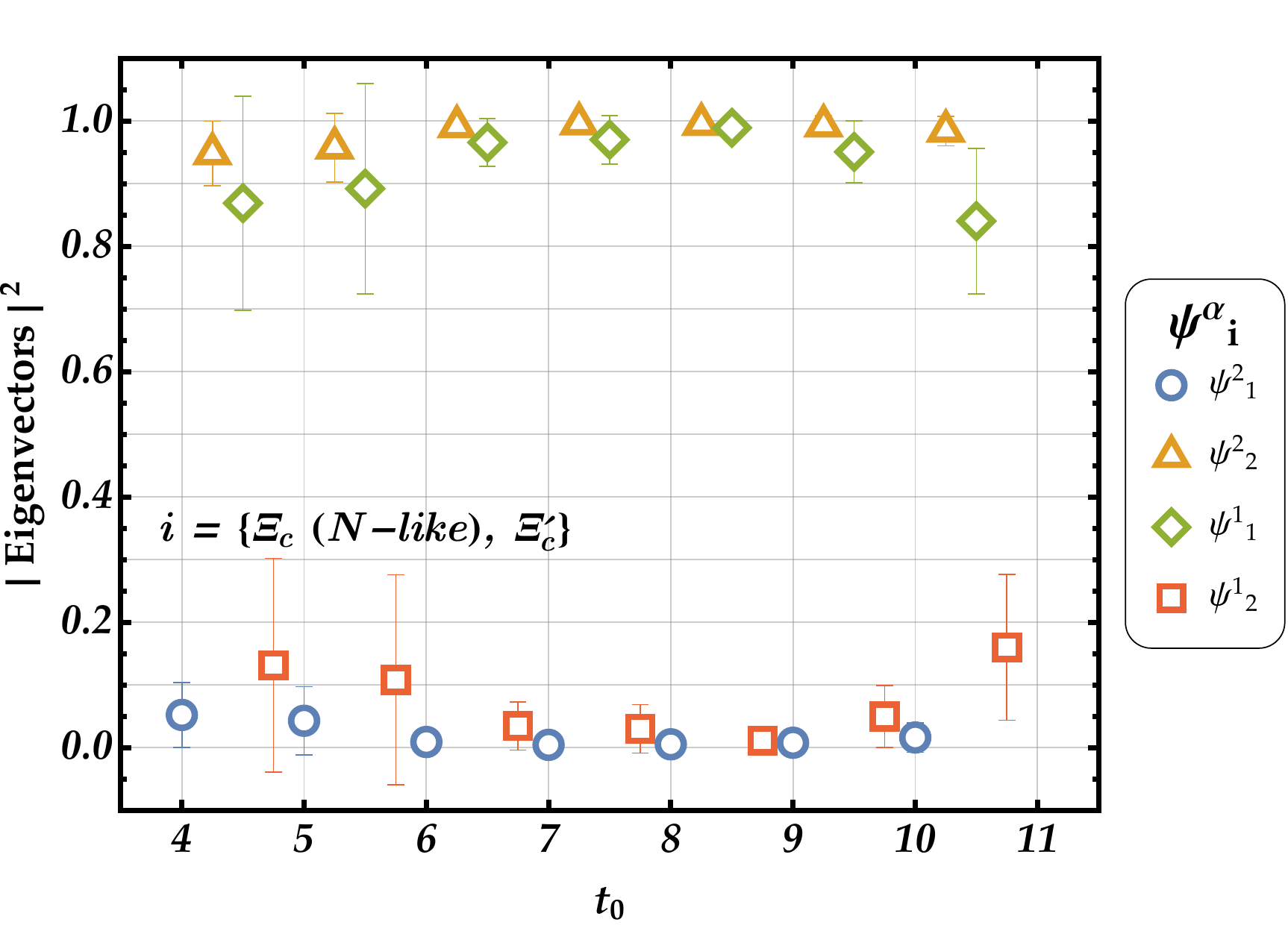} }
						\caption{ \label{fig:xiccross_opComp1} Ground state signals for the $\Xi_c \left( \frac{1}{2}^\pm \right)$ and $\Xi_c^\prime \left( \frac{1}{2}^\pm \right)$ channels, the former obtained via $\Lambda$-like and $N$-like operators. The left (right) three panels show the positive (negative) parity results. Eigenvectors for variational solutions of the $i=\{\Xi_c (\Lambda\text{-like}), \Xi_c^\prime\}$ (filled) and $i=\{\Xi_c (N\text{-like}), \Xi_c^\prime\}$ (hollow) operator sets for both positive and negative parity channels are given to show the strength of the mixing between operators. The state index $\alpha$ follows the order of $i$ and is directly related to the signals in the upper effective mass plots. For instance, the filled green diamond $\psi^1_1$ in the lower leftmost eigenvector plot indicates the $\Xi_c (\Lambda\text{-like})$ signal associated with the $\Xi_c (\Lambda\text{-like})$ operator. The hollow red square $\psi^1_2$ of the lower rightmost plot is the $\Xi_c^\prime$ contribution to the $\Xi_c (N\text{-like})$ signal. }
					\end{figure*}

					\paragraph{$\Lambda$-like operators:} $\Lambda_c$ and $\Xi_c$ belong to the totally flavor-antisymmetric $SU(4)$ anti-quadruplet and hence are studied via the flavor-octet $\Lambda$-like operators. The behaviors of these operators depicted in \Cref{fig:opComp_xic12p,fig:opEvecs} show similarities to the $N$-like $\Xi_{cc}$ case. It can naively be expected that the first term of the $\Lambda$-like operator (see \Cref{tab:intop2}) would have the dominant contribution, which would mean that it is in essence the same as the $N$-like operator. Indeed, by rearranging the latter two terms of the $\Lambda$-like operator via Fierz transformations, one can show that the coefficient of the $\left[ q_1^{Ta}(x) C \gamma_5 q_2^b(x) \right] q_3^c(x)$ term of the operator is five times the other resulting terms. The same argument holds for the other Dirac structures as well. This dominance is realized in our comparisons of the $\Xi_c(\frac{1}{2}^+)$ results illustrated in \Cref{fig:xiccross_opComp1}, where we have an almost identical signal for the ground states calculated via the $\Lambda$-like and the $N$-like operator. 

					Additionally, the flavor decomposition of the $\Lambda_c$ studied in Ref.~\cite{Gubler:2016viv} by three of the present authors shows that the negative parity $\Lambda_c$ baryon consists of a mixture of flavor-singlet and flavor-octet wave-functions. The flavor-octet interpolating operator that we employ for the $\Lambda_c$ baryon may therefore be inadequate to resolve the lowest-lying negative parity state by itself. A similar conclusion was reached in Ref.~\cite{PhysRevLett.108.112001}. The first excited negative parity state on the other hand, is dominated by a flavor-octet wave-function and it is possible that this state is contaminating our lowest $\Lambda_c (\frac{1}{2}^-)$ signal, which could be a plausible explanation of the apparent overestimation of its mass (see \Cref{tab:barmass} and \Cref{fig:comp}).    

					We analyze the $\Xi_c$ channel with two different types of operators. One being the $\Lambda$-like, the other the $N$-like operator as given in \Cref{tab:intop2}. We find that both give consistent results for the positive parity case while there is a difference for negative parity. As shown in \Cref{fig:opComp_xic12n}, the $N$-like operator couples to a lower-lying state for the $\{\chi_1, \chi_4 \}$ basis. Similar differences between these operators for the negative parity sector have been reported by the RQCD Collaboration~\cite{PhysRevD.92.034504}.

					\paragraph{$\Xi_c - \Xi_c^\prime$ mixing:} We perform a correlation matrix analysis consisting of the $\Xi_c^\prime$, and $N$-like and $\Lambda$-like $\Xi_c$ operators in order to investigate the possible mixing between these baryons. We construct the correlation-function matrices for this analysis in two steps. First, we solve a variational system over the $\{\chi_1, \chi_4 \}$ basis for each element of the correlation matrix and take the lowest lying state. We find that this approach helps to isolate the ground states better. We then solve another $2 \times 2$ correlation matrix with both $\Xi_c$ and $\Xi_c^\prime$ ground state operators to investigate the mixing effects. 

					For positive parity $\Xi_c$ and $\Xi_c^\prime$, we analyze the cross correlators between the flavor-octet $SU(4)$ $\Xi_c$ - $\Xi_c^\prime$, and the $N$-like $\Xi_c$ - $\Xi_c^\prime$ individually. We find that the $\Xi_c$ and $\Xi_c^\prime$ signals separate nicely, and the $N$-like and $\Lambda$-like $\Xi_c$ operators produce consistent signals with negligible mixing (see \Cref{fig:xiccross_opComp1}). Magnitudes of the eigenvectors also confirm that the $\Xi_c$ and $\Xi_c^\prime$ states have distinct signals. In case of negative parity, there appears to be non-negligible mixing between the two states dependent on the variational parameters. Specifically, the $\Lambda$-like $\Xi_c$ has a negligible $\Xi_c^\prime$ component, while the $N$-like $\Xi_c$ state has up to a $10\%$ $\Xi_c^\prime$ mixing although the effect seems to depend on the variational parameters. The reason why the negative parity $\Lambda$-like operator gives signals close to the $\Xi_c^\prime$ is understood to be related to the overestimation of the mass obtained for that operator rather than a mixing effect. The $\Xi_c^\prime$ appears to have at most a mixing of $5\%$ with the $N$-like $\Xi_c$. In all, we see that for negative parity the mixing is not completely negligible, but nevertheless quite small. 

					We note that the quantitative analysis given here should not be considered as the definitive mixing between the $\Xi_c$ and the $\Xi_c^\prime$ states but rather the mixing between the operators that we utilize in this work. Overlap factors of the correlation functions and the eigenvectors are dependent on the smearing of the quark fields. Hence the amount of mixing differs for different smearing parameters.     

				\subsubsection{Smearing dependence}\label{sec:opSmr}
					\paragraph{Spin-$1/2$ baryons:} We observe that, evidently, the ground state signals remain stable with respect to the smearing radius. The excited-state signals on the other hand show a clear dependence to the smearing radius of the source quark fields. This is readily visible for every case given in \Cref{fig:opComp}. For both positive and negative parity, states that are clearly separated from the ground state tend to decrease as the smearing radius increases with no apparent plateau behavior. Note that all the energies are extracted via a plateau approach, which are dependent on the choice of the fit windows. Extracting the energies from two-exponential fits are more reliable for the $\sim 0.2$ and $0.4$ smearings, where those fit results coincide with that of $\sim 0.7$ extracted via a plateau approach or a two-exponential fit. This indicates that the signals of the widest smearing are the most reliable to estimate the energy levels.

					When we enlarge the operator basis by combining two operators with two different smearings and perform a $4 \times 4$ analysis, we end up with quite noisy solutions due to the current limited statistics, which renders a conclusive analysis impossible. We, however, observe an apparent degeneracy in three out of four solutions as shown in the $\{\chi_1,\chi_4\}-\{10,50\}$ row of \Cref{fig:opComp_xicc12p}. A similar behavior is seen for other combinations of operators and smearings as well. Investigations of the eigenvectors show that all states, \textit{e.g.} ground or excited states, couple to the operators with the wider quark sources. This is confirmed independently if we compare the higher states in the rows $\{\chi_1,\chi_4\}-50$ and $\{\chi_1,\chi_4\}-\{10,50\}$ of \Cref{fig:opComp_xicc12p}, where the extracted values coincide. We find this to be true for any variational analysis over multiple smearings.         

					\paragraph{Spin-$3/2$ baryons:} We find that solving a $3 \times 3$ variational system with smeared-smeared operators only, provides no additional information compared to a $2 \times 2$ system with the smearings at hand. One solution turns out to be indistinguishable from the other so we focus on the solutions from the two narrower smearings, which give less noisier signals. 

			\subsection{Charmed baryon spectrum}\label{sec:cspec}
				The energy levels from the diagonalized correlation functions are extracted by fitting the data to the form given in \Cref{eq:dc}. Additional exponential terms are employed to stabilize the fits against excited-state contributions. In most of the cases, where the signal forms a plateau in the effective-mass plots, masses of the lowest states extracted from the one-exponential fits agree with the multi-exponential fit results within their error bars. Yet, a two-exponential form stabilizes the fits and improves the accuracy of the results. This is especially true when analyzing the widest smearing case. The extracted energies are compiled in \Cref{tab:barmass}. Since we are at the isospin-symmetric point, $m_u = m_d$, our results should be understood as the isospin averaged masses of the respective states. 
				\begin{table*}[htbp]
					\caption{ \label{tab:barmass} Extracted baryon masses in units of GeV. }
					\renewcommand{\arraystretch}{1.4}
					\centering
					\begin{tabularx}{\textwidth}{L{1}L{1}C{1}C{1}L{1}C{1}C{1}C{1}}
						\hline\hline
						Baryon 			& $J^P$ 			& $M_1$ 	& $M_2$ 
										& $J^P$ 			& $M_1$ 	& $M_2$ 	 & $M_3$ \\
						\hline	
						$\Lambda_c$ 	& $\frac{1}{2}^+$	& 2.343(23) & 3.280(76)  
										& $\frac{1}{2}^-$ 	& 2.668(16) & 2.992(14)  & 3.439(29) \\
						$\Sigma_c$ 		& $\frac{1}{2}^+$	& 2.459(45) & 3.270(68)  
										& $\frac{1}{2}^-$	& 2.814(20) & 2.854(17)  & 3.541(45) \\
						$\Xi_c$ 		& $\frac{1}{2}^+$ ($\Lambda$-like)	& 2.474(11) & 3.301(33)  
										& $\frac{1}{2}^-$ ($N$-like)	& 2.770(67) & 3.059(10)  & 3.390(76) \\
						$\Xi_c^\prime$ 	& $\frac{1}{2}^+$	& 2.593(22) & 3.072(45)  
										& $\frac{1}{2}^-$	& 2.933(16) & 2.968(19)  & 3.338(88) \\
						$\Omega_c$ 		& $\frac{1}{2}^+$	& 2.711(16) & 3.392(11)  
										& $\frac{1}{2}^-$	& 3.044(15) & 3.069(17)  & --- \\	
						$\Sigma_c^\ast$ & $\frac{3}{2}^+$	& 2.508(45) & 3.648(23)  
										& $\frac{3}{2}^-$	& 2.797(38) & 4.428(40)  & --- \\										
						$\Xi_c^\ast$ 	& $\frac{3}{2}^+$	& 2.648(37) & 3.637(202) 
										& $\frac{3}{2}^-$	& 2.803(135)& --- 		 & --- \\
						$\Omega_c^\ast$ & $\frac{3}{2}^+$	& 2.773(16) & 3.449(167) 
										& $\frac{3}{2}^-$	& 3.066(43) & --- 		 & --- \\
										
						$\Xi_{cc}$ 		& $\frac{1}{2}^+$	& 3.615(33) & 4.327(47)  
										& $\frac{1}{2}^-$	& 3.930(20) & 3.971(22)  & 4.246(193) \\
						$\Omega_{cc}$ 	& $\frac{1}{2}^+$	& 3.733(13) & 4.417(32)   
										& $\frac{1}{2}^-$	& 4.041(15)	& 4.063(13)  & 4.395(41) \\										
						$\Xi_{cc}^\ast$ & $\frac{3}{2}^+$	& 3.703(33) & 4.129(40)  
										& $\frac{3}{2}^-$	& 4.009(31) & --- 		 & --- \\
					$\Omega_{cc}^\ast$ 	& $\frac{3}{2}^+$	& 3.793(30) & 4.196(97)  
										& $\frac{3}{2}^-$	& 4.115(70) & --- 		 & --- \\
										
						$\Omega_{ccc}$ 	& $\frac{3}{2}^+$	& 4.817(12) & 5.417(40)  
										& $\frac{3}{2}^-$	& 5.083(67) & 5.734(97)  & --- \\
						\hline\hline
					\end{tabularx}
				\end{table*}

\begin{figure*}[!htb]
					\centering
					\includegraphics[width=.49\textwidth]{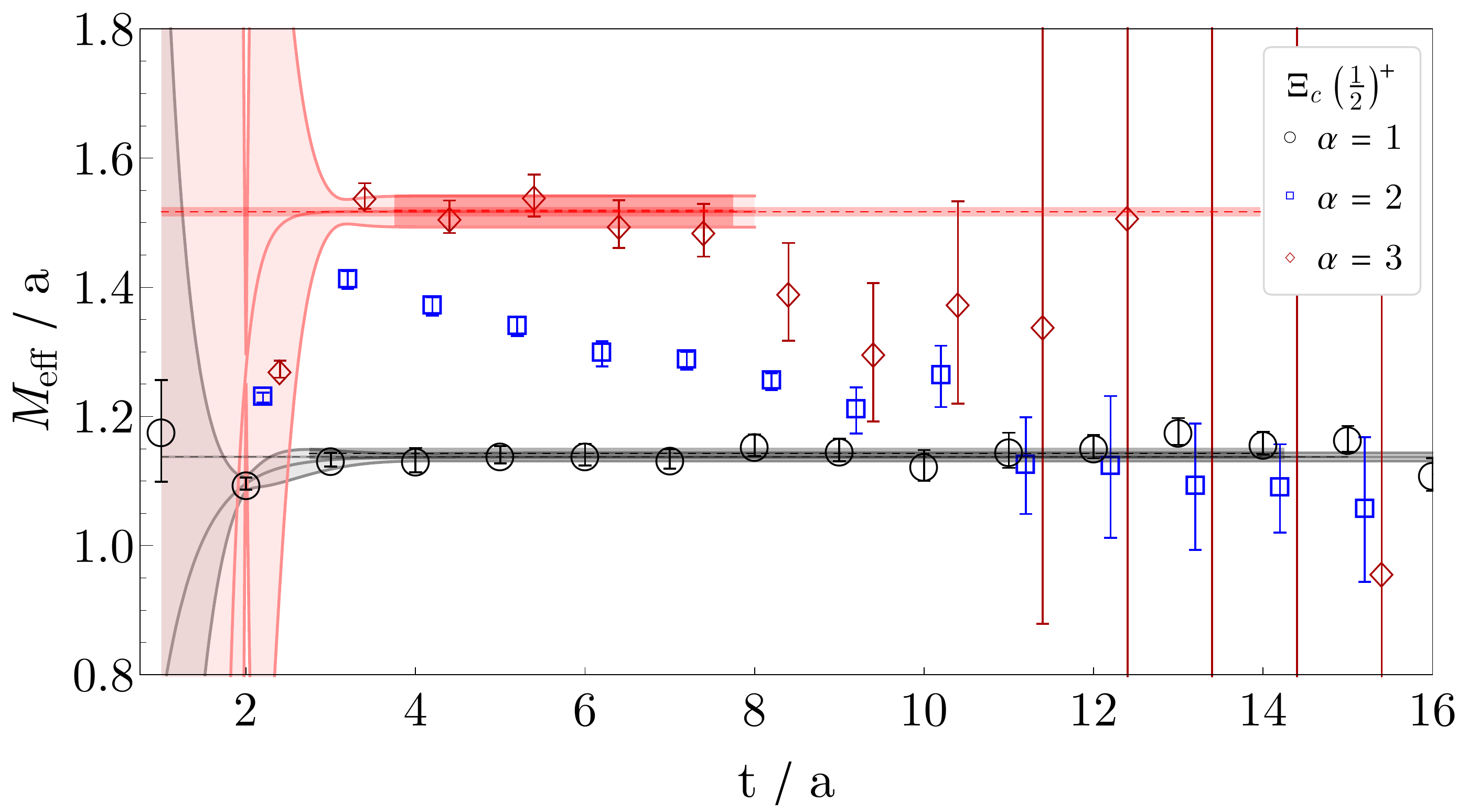} 
					\includegraphics[width=.49\textwidth]{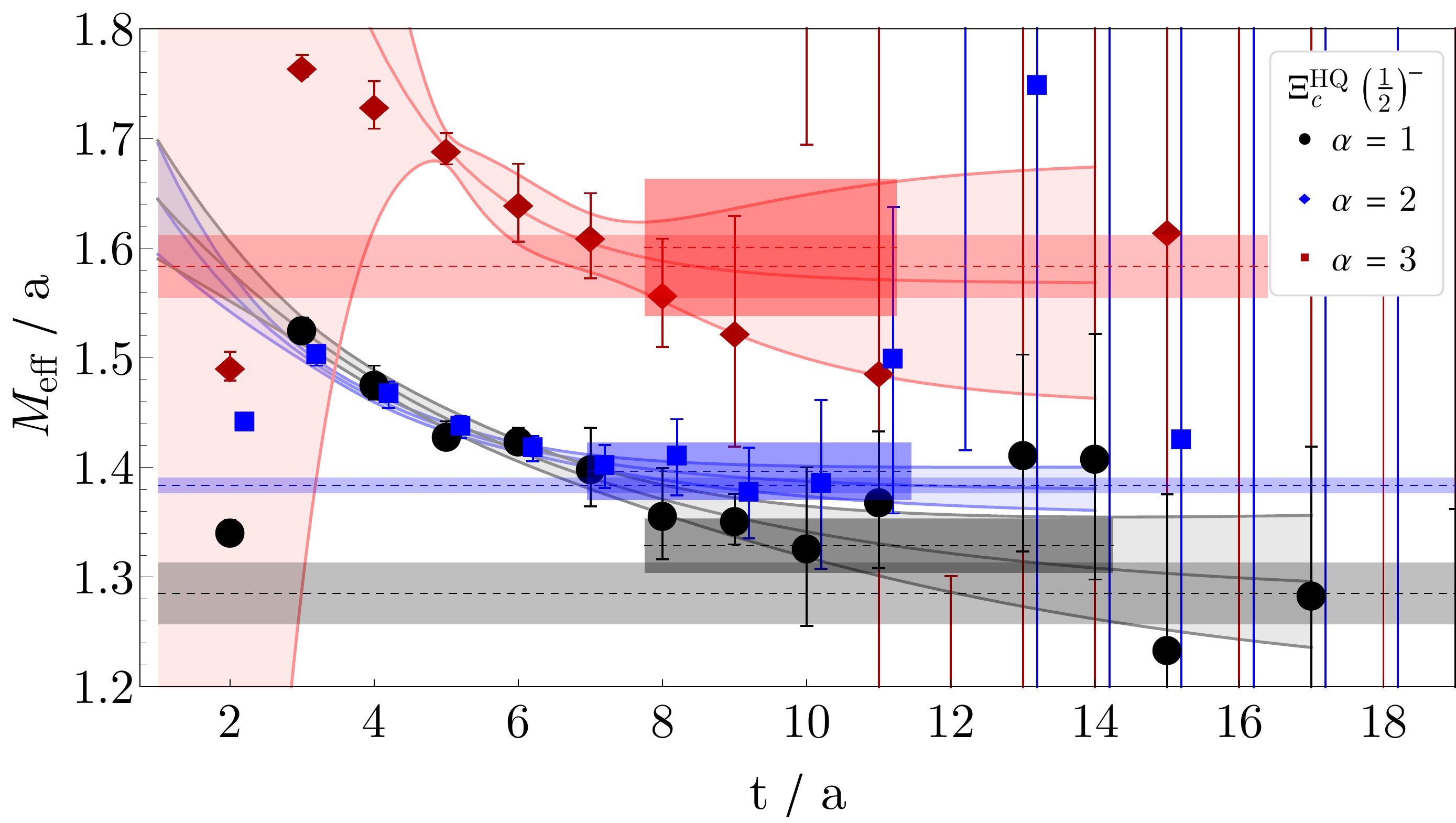} \\
					\includegraphics[width=.49\textwidth]{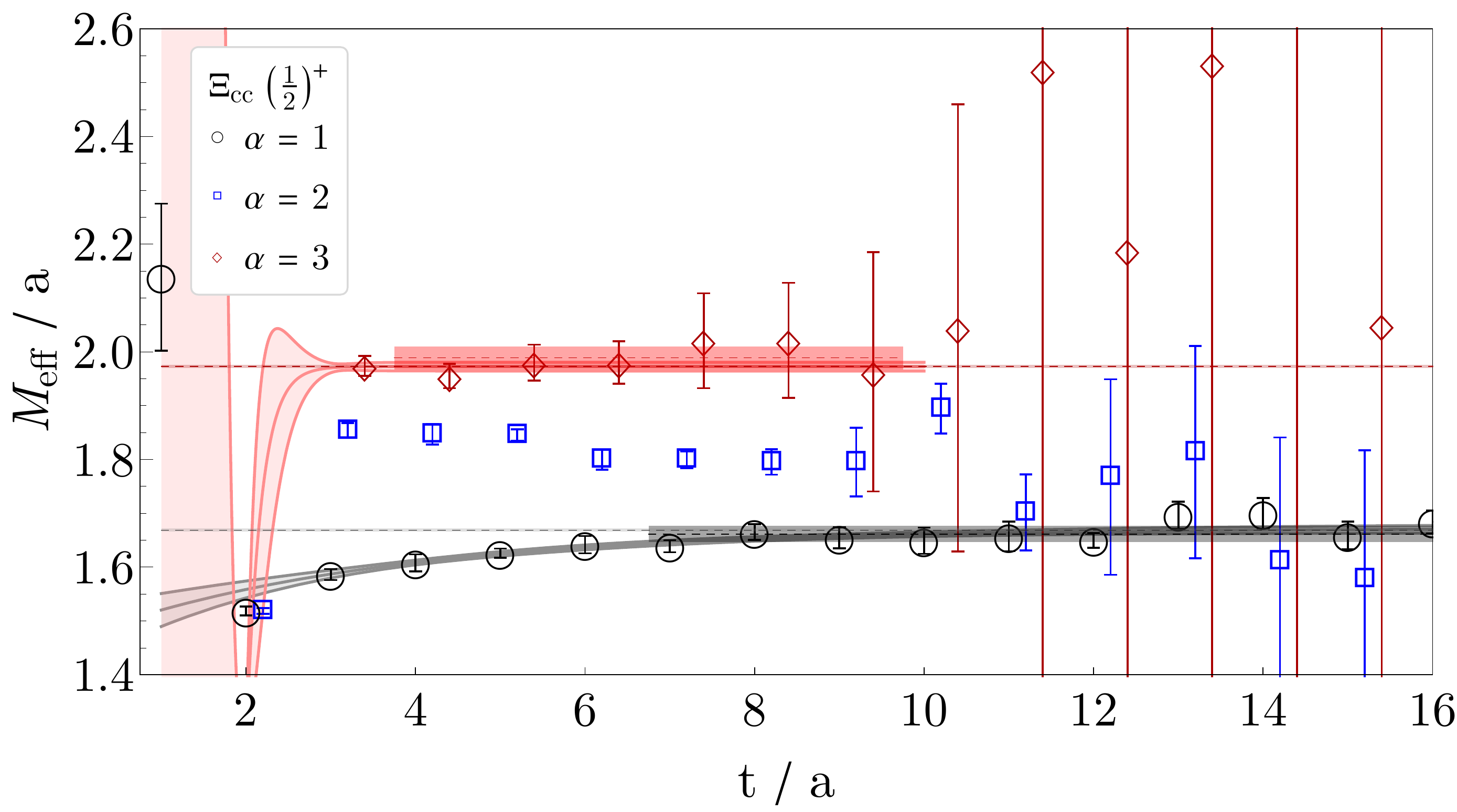} 
					\includegraphics[width=.49\textwidth]{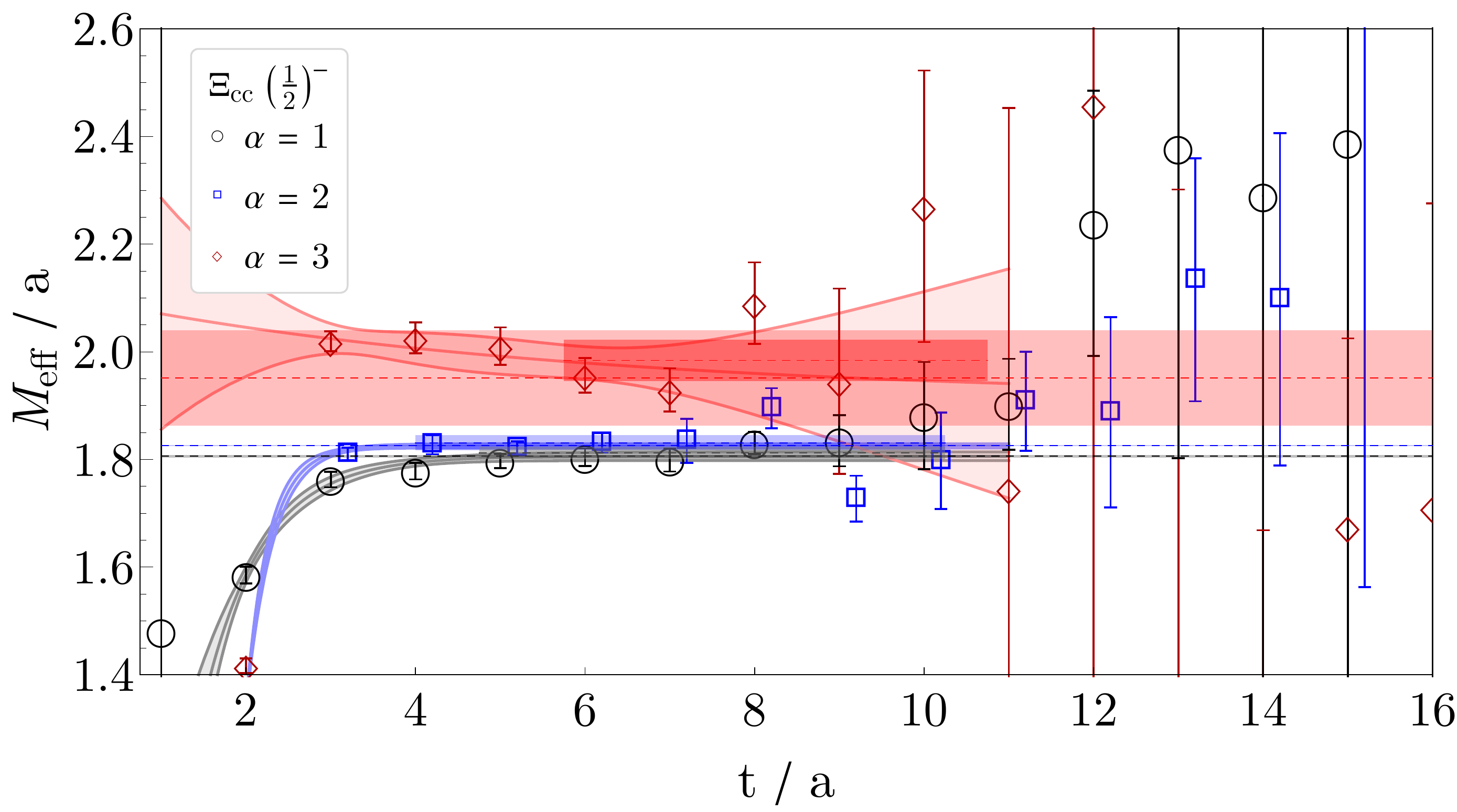} \\
					\includegraphics[width=.49\textwidth]{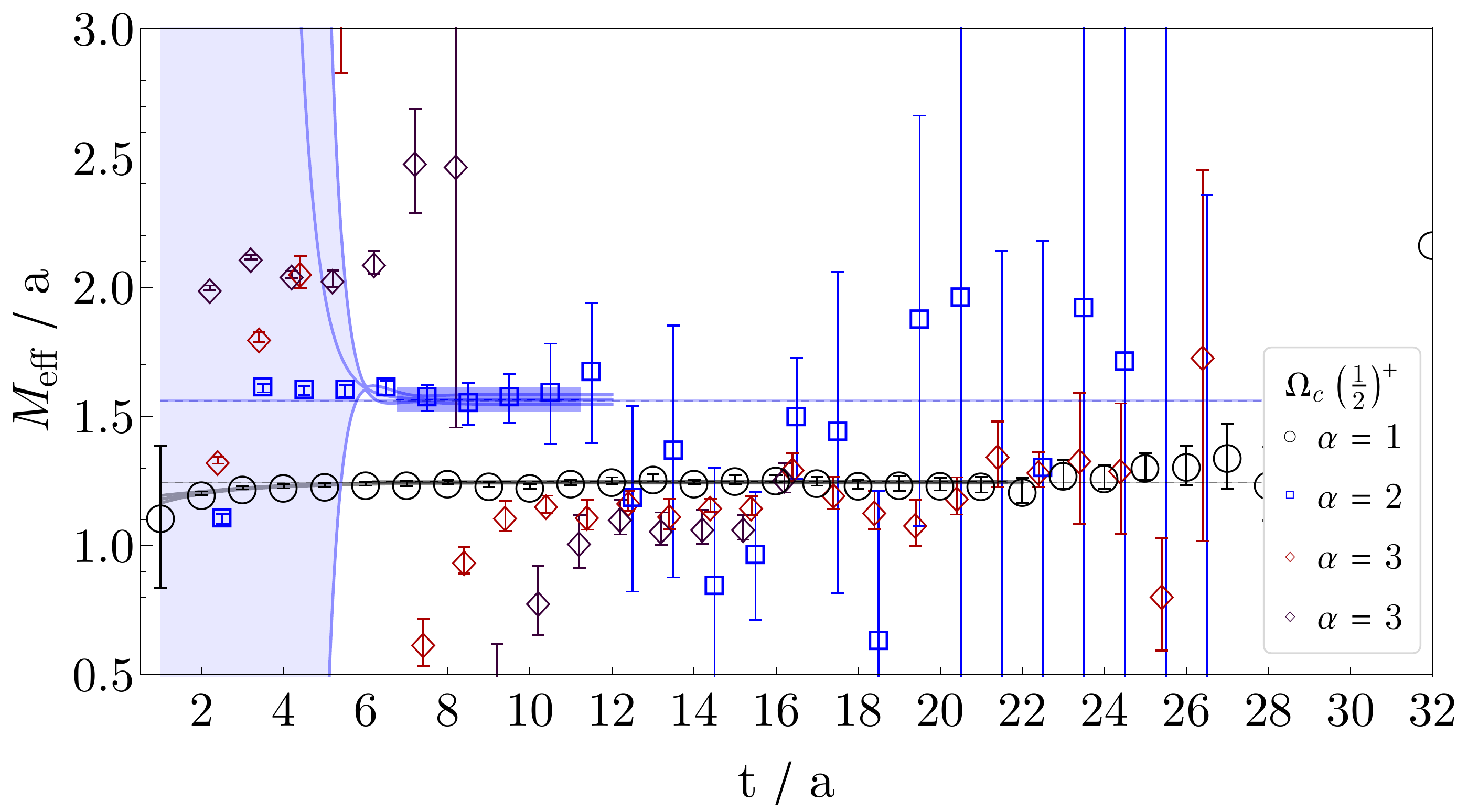} 
					\includegraphics[width=.49\textwidth]{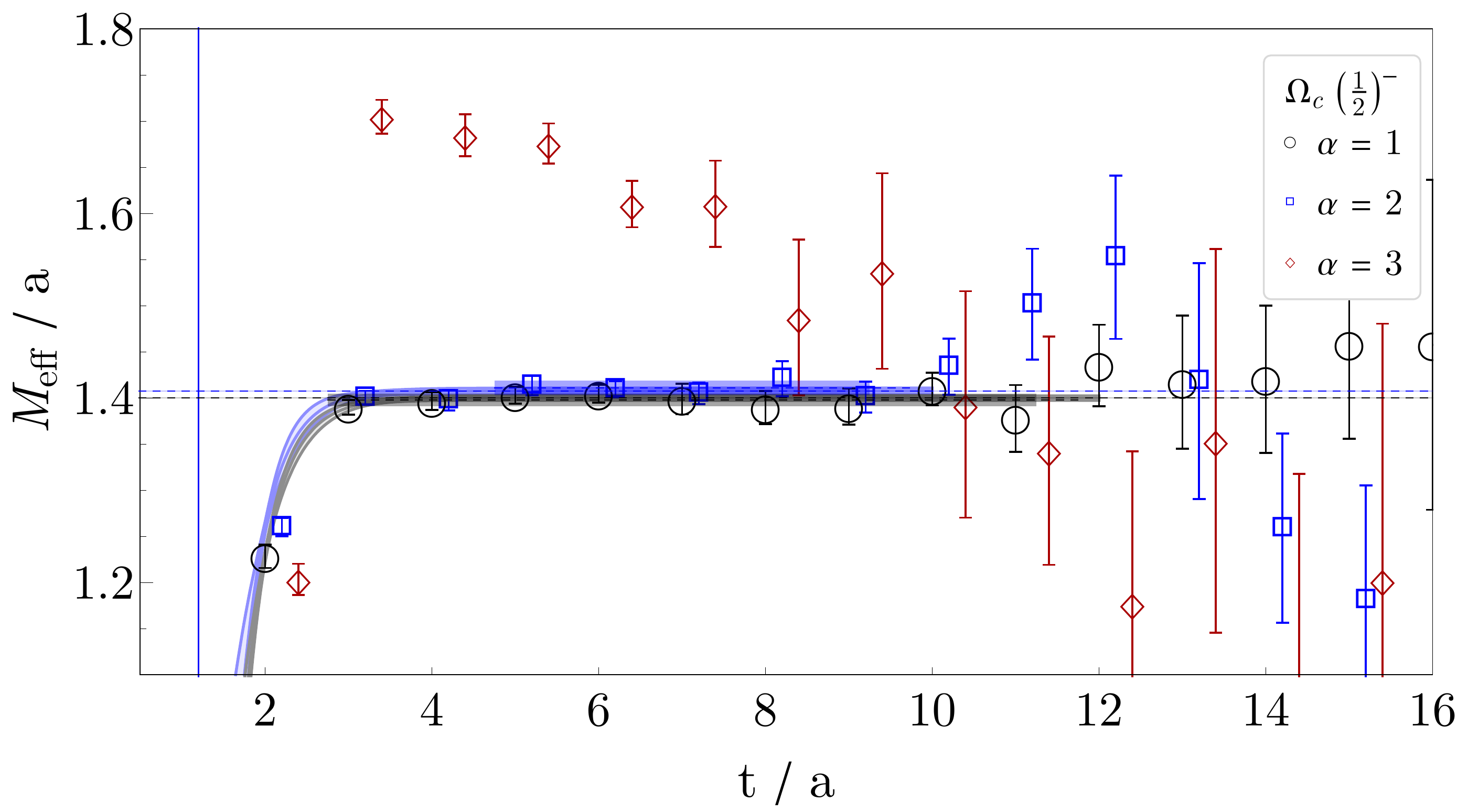}
					\caption{ \label{fig:effmasses} Effective mass plots for representative baryons. Colored curves show the weighted two-exponential fits to the central points. Bands spanning the plots show the energy levels and their 1$\sigma$ uncertainties extracted via configuration-by-configuration two-exponential fits. Plateau approach fit windows (colored rectangles) are shown for comparison only, as the two-exponential fit is our preferred method of choice. We note that, although there appears a ``bump'' around $t/a=10$, blue data points of the $\Xi_c(1/2^+)$ and $\Xi_{cc}(1/2^+)$ plots show a decreasing trend (see the discussion in \Cref{sec:cspec}). The $\Omega_c(1/2^+)$ plot shows the signals that correspond to the narrowest and the widest smearings of the third state in relation to the discussion in \Cref{sec:cspec}.}
				\end{figure*}

				\paragraph{Operator basis and the quality of the signals:}
				As we have discussed in \Cref{sec:opSmr}, a variational analysis over a set of different smearings for a fixed operator returns solution eigenvectors that couple to the widest smearing. Therefore, we always use an operator basis with quark smearings fixed to the widest one. For  the spin-$1/2$ cases, we perform $3 \times 3$ variational analyses with a fixed smearing over the operator sets $\{\chi_1, \chi_2, \chi_4 \}$ and extract signals of three states for each channel. The third energy level with largest energy is however usually lost to noise already at relatively early time slices or decays to the ground states due to inaccuracies in the diagonalization procedure of \Cref{eq:gev,eq:dc}. For instance, in case of the positive parity spin-$1/2$ $\Xi_c$ baryons, we find that the state dominantly coupling to the $\chi_2$ operator decays to the ground state signal before showing a plateau that may be a candidate signal for an excited state (blue rectangles in the top left plot of \Cref{fig:effmasses}). Signals of possible third states for the spin-$1/2$, positive parity $\Sigma_c$, $\Xi_c^\prime$ and $\Omega_c$ channels emerge in early time slices of effective mass analyses but are quickly lost to noise. It is usually possible to identify a fit region of 2-3 points for the narrowest smearing but we find the energy extracted via this approach to be unreliable, since the fit window is very small and the smearing dependency of the state cannot be established. Positive parity spin-$1/2$ $\Xi_{cc}$ and $\Omega_{cc}$ signals mimic the behavior of $\Xi_c$, where there appear signals one could potentially identify as distinct states. However we find that those states are rather unstable under the change of variational parameters. In addition, extracted energies are highly dependent on the extraction method -- plateau approach or a two-exponential fit. Therefore, even though we show their signals in the plots, we do not extract or report any corresponding energy values.

				In general, we find that the negative parity sector appears to be richer in comparison to the positive parity case. Indeed, we could identify three distinct states for most of the negative parity spin-$1/2$ channels. Isolating the low-lying states via a plateau approach is a challenge here since 
				multiple energy levels appear in a narrow energy range. Two-exponential fits are very helpful in such cases to disentangle and extract the states more accurately. A relatively early time slice is needed for the two-exponential fits to perform effectively. We choose the initial time slices from the range $t_i = [2,5]$. No significant dependence to this choice is observed for the ground and first excited states. Second excited states in the negative parity sector are relatively more susceptible to the choice of the initial time slice, however. The systematic uncertainties associated with those are less than the current statistical errors on their extracted energies.  Effective mass plots illustrating the above discussions are given in \Cref{fig:effmasses}.

				\paragraph{Mass differences:} Hyperfine splittings, the mass differences between the spin-$3/2$ and spin-$1/2$ states, of the $\Sigma_c$, $\Xi_c$, and $\Omega_c$ channels are reproduced in good agreement with the experimental values. Mass differences between the positive and negative parity states also agree well with the available experimental results. The first excited states of the positive parity baryons lie quite high, $400$ MeV to $1$ GeV, above the ground states. A common pattern is that, more than one negative parity state for the singly- and doubly-charmed spin-$1/2$ baryons appear in between the positive parity ground and first-excited state. The first two negative parity states of the $\Sigma_c$, $\Xi_c^\prime$, $\Omega_c$, $\Xi_{cc}$, and $\Omega_{cc}$ channels lie close to each other. The splittings between those states are smaller for the $\Omega_c$ and $\Omega_{cc}$ baryons compared to those of $\Sigma_c$, $\Xi_{c}^\prime$, and $\Xi_{cc}$. The situation is different for the $\Lambda_c$ and the $\Xi_c$ baryons where the negative parity states are roughly $300$ MeV apart.

\begin{figure*}[!htb]
					\centering
					\includegraphics[width=.49\textwidth]{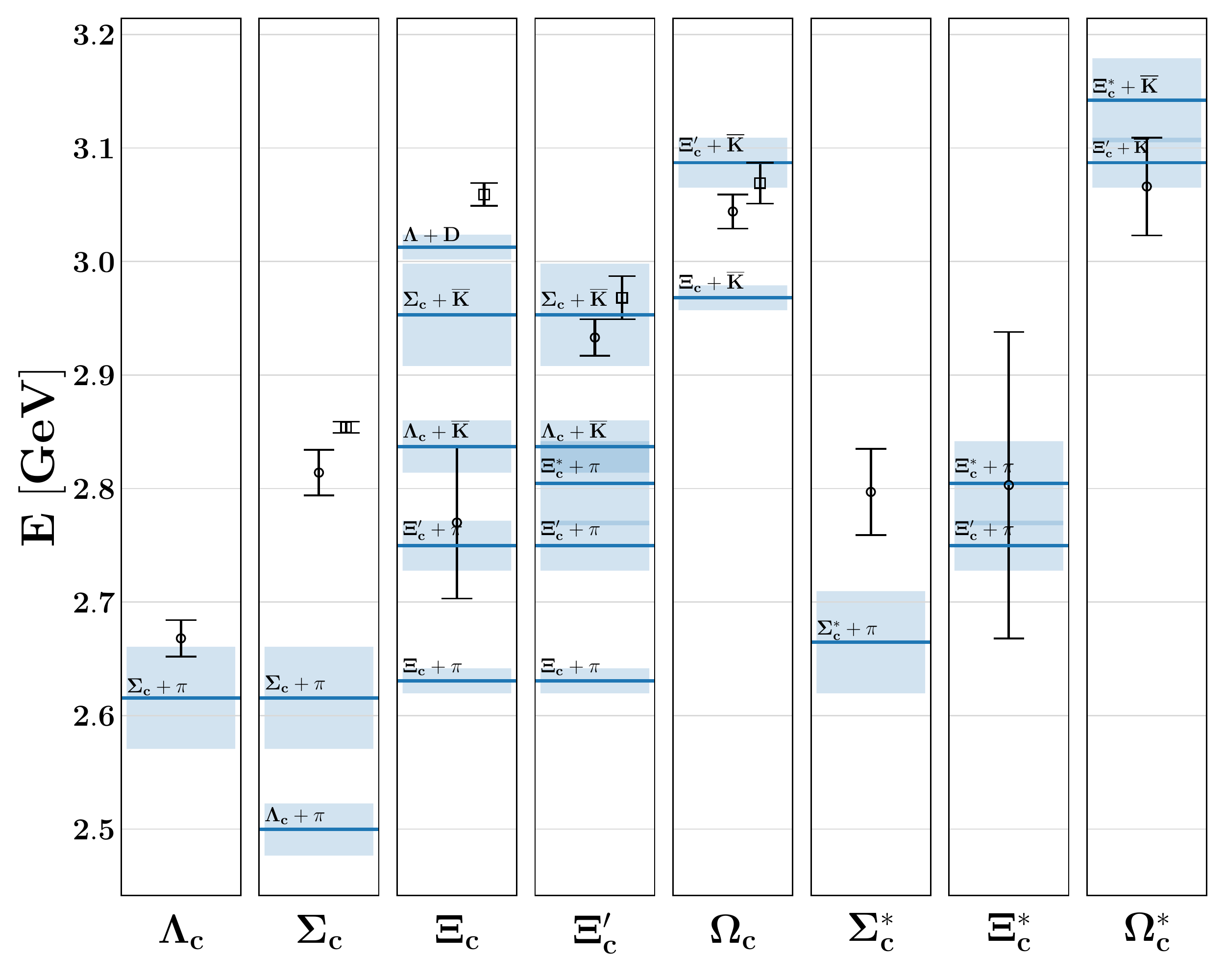}
					\includegraphics[width=.245\textwidth]{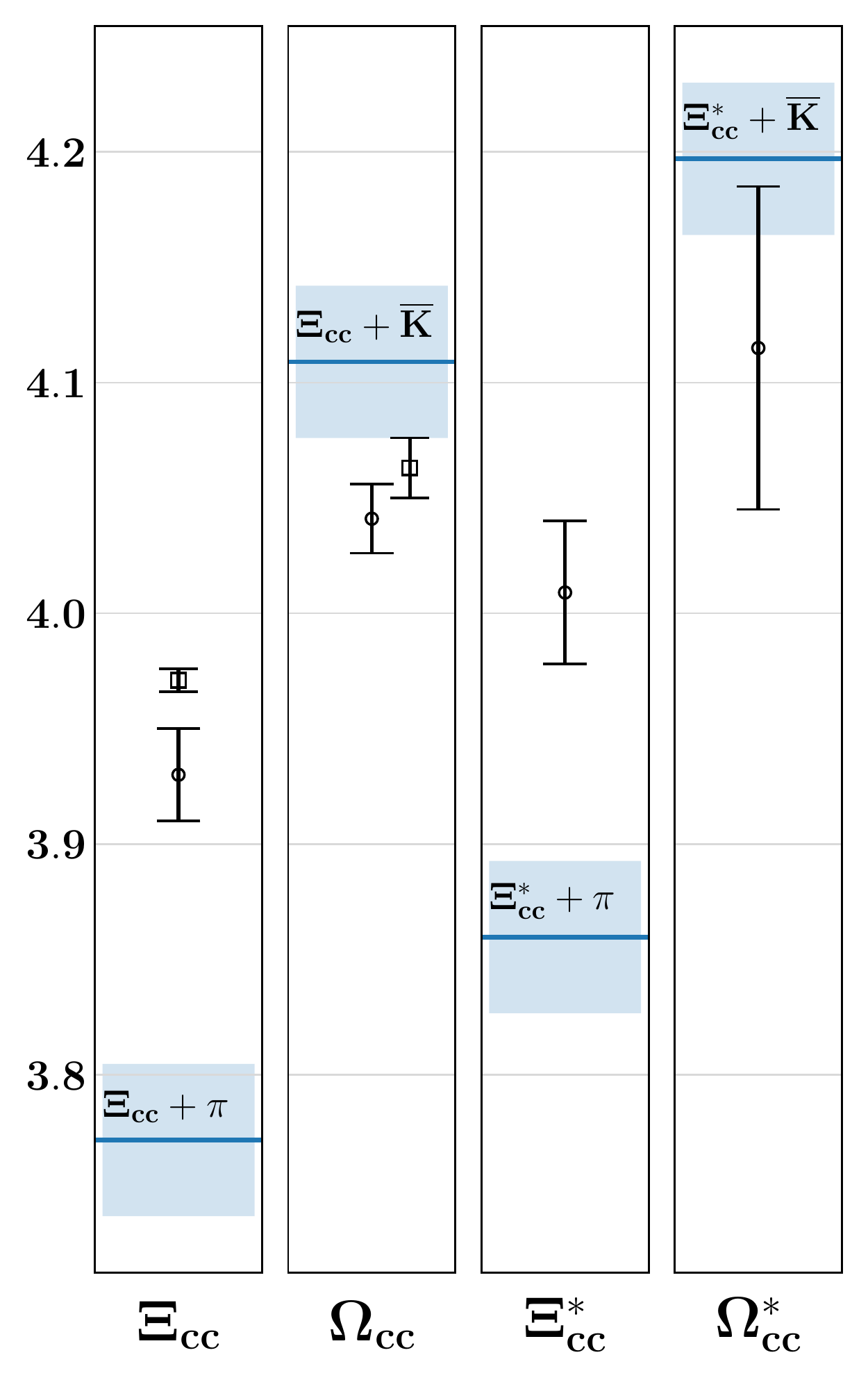}
					\caption{ \label{fig:threshold_comp} The $S$-wave scattering thresholds for each charmed baryon channel. Open symbols are the extracted energies of the negative parity baryons, given in \Cref{tab:barmass}, that lie close to the thresholds. Horizontal lines with the shaded regions are the calculated threshold energies with the statistical errors associated with the baryon energies' uncertainties only. See the \emph{Scattering states} part of \Cref{sec:cspec} for our treatment of the $\Lambda + D$ threshold. }
				\end{figure*}

				\paragraph{Scattering states:} It is essential to examine the relevant thresholds for the negative parity states in order to check if they could correspond to scattering states. It is possible for the negative parity ground states to couple to the $S$- or $D$-wave scattering states of a positive parity baryon and a negative parity meson. The relevant thresholds which respect to isospin, spin, parity, strangeness and charm quantum numbers are,
				\begin{align*}
					\Lambda_c &\rightarrow \Sigma_c + \pi, &
					\Sigma_c  &\rightarrow \Lambda_c + \pi, &
					\Sigma_c &\rightarrow \Sigma_c + \pi, \\
					\Sigma_c^\ast &\rightarrow \Sigma_c^\ast + \pi, &
					\Xi_c &\rightarrow \Xi_c + \pi, &
					\Xi_c &\rightarrow \Xi_c^\prime + \pi, \\
					\Xi_c &\rightarrow \Xi_c^\ast + \pi, &
					\Xi_c &\rightarrow \Lambda_c + \overline{K}, &
					\Xi_c &\rightarrow \Sigma_c + \overline{K},\\
					\Xi_c &\rightarrow \Lambda + D, &
					\Xi_c^\prime &\rightarrow \Xi_c + \pi, &
					\Xi_c^\prime &\rightarrow \Xi_c^\prime + \pi, \\
					\Xi_c^\prime &\rightarrow \Xi_c^\ast + \pi, &
					\Xi_c^\prime &\rightarrow \Lambda_c + \overline{K}, &
					\Xi_c^\prime &\rightarrow \Sigma_c + \overline{K}, \\
					\Xi_c^\ast &\rightarrow \Xi_c^\ast + \pi, &
					\Omega_c &\rightarrow \Xi_{c} + \overline{K}, &
					\Omega_c &\rightarrow \Xi_{c}^\prime + \overline{K}, \\
					\Omega_c^\ast &\rightarrow \Xi_{c}^\ast + \overline{K}, &
					\Xi_{cc} &\rightarrow \Xi_{cc} + \pi, &
					\Xi_{cc}^\ast &\rightarrow \Xi_{cc}^\ast + \pi, \\
					\Omega_{cc} &\rightarrow \Xi_{cc} + \overline{K}, &
					\Omega_{cc}^\ast &\rightarrow \Xi_{cc}^\ast + \overline{K}.
				\end{align*}
				We plot the above two-particle thresholds together with the extracted negative parity energies in \Cref{fig:threshold_comp}. The two-particle scattering energies are calculated via $E = \sqrt{M_1^2 + \mathbf{p}_1^2} + \sqrt{M_2^2 + \mathbf{p}_2^2}$, where $M_i$ is the mass of the particle and $\mathbf{p}_i = 2\pi\mathbf{n}/L$ the lattice momentum. We use the $\pi$ mass quoted in the PACS-CS paper~\cite{PhysRevD.79.034503} and the experimental $K$ mass, since we use a strange quark mass re-tuned to its physical value via the $K$ mass input~\cite{PhysRevLett.108.112001}, along with the positive parity baryon masses from \Cref{tab:barmass} of this work in calculating the threshold energies. The $\Lambda + D$ threshold has to be estimated differently since we do not calculate the $\Lambda$ baryon or the $D$ meson in this work. In estimating the threshold, we take the experimental $\Lambda$ mass and multiply it by a correction factor, $\Lambda_c^{our}/\Lambda_c^{exp}$, due to our overestimation of the $\Lambda_c$ mass. The uncertainty of this value is assumed to be same as that of $\Lambda^{our}_c$. The $D$ meson mass is taken to be its experimental value with its uncertainty neglected. The momenta $\mathbf{p}_1$ and $\mathbf{p}_2$ are set to zero. An inspection of \Cref{fig:threshold_comp} shows that some of the $\Xi_c$ baryon signals may contain scattering states because of their vicinity to various thresholds. Indeed, $M_1[\Xi_c(\frac{1}{2}^-)]$, $M_{1,2}[\Xi_c^\prime(\frac{1}{2}^-)]$, and $M_1[\Xi_c^\ast(\frac{3}{2}^-)]$ lie close to at least one related threshold. We also find some states that lie above the thresholds to be close to their respective boosted ($\mathbf{n} > 0$) thresholds.

				\paragraph{Negative parity $\Sigma_c$ states:} Three of our negative parity $\Sigma_c$ states lie close to the PDG-listed $\Sigma_c(2800)$ baryon. BABAR reports a direct mass measurement of the $\Sigma_c^0$ state as $M[\Sigma_c^0(2800)]=2846 \pm 18$ MeV/c$^2$. Belle, on the other hand, identifies the $\Sigma_c(2800)$ state from the signals seen in the distribution of the mass difference, $\Delta M(\Lambda_c^+ \pi) \equiv M(\Lambda_c^+ \pi) - M(\Lambda_c^+)$. The corresponding $\Sigma_c^0(2800)$ mass reported in the PDG based on this measured difference is $M[\Sigma_c^0(2800)]=2806^{+5}_{-7}$ MeV/c$^2$, 40 MeV/c$^2$ lower than that of BABAR. It is noted in the PDG listings that the state that has been observed by BABAR might be a different $\Sigma_c$ excitation. 

				Given that these states have been seen in the $\Lambda_c \pi$ invariant mass spectra, a straightforward assignment for the quantum numbers would be $J^P=1/2^-$. From a quark model perspective (see paragraph $g.$), there are three possible low-lying negative parity spin-1/2 $\Sigma_c$ excitations. Two $\lambda$-modes with diquark spin $j=0$ and $j=1$, and a $\rho$-mode with diquark spin $j=1$. In the heavy quark limit, the $S$-wave $\Sigma_c(2800) \to \Lambda_c \pi$ transitions of the $j=1$ $\lambda$- and $\rho$-modes would be forbidden due to the violation of the spin-parity conservation of the light-quark degrees of freedom. A heavy quark effective theory calculation estimates a very large decay width, of the order of 885 MeV, for the $j=0$ $\lambda$-mode~\cite{Chengs:2015hy}, which rules out the $1/2^-$ quantum number for $\Sigma_c(2800)$. On the other hand, a $D$-wave transition is possible and points to the $J^P=3/2^-, 5/2^-$ possibilities. The lowest-lying $\Sigma_c^\ast(\frac{3}{2}^-)$ state we extract with a mass of $M_1 = 2797 \pm 38$ MeV/c$^2$, might therefore be a better suited candidate, which is situated in the vicinity of the masses $M[\Sigma_c^{++}(2800)]=2801^{+4}_{-6}$ MeV/c$^2$ and $M[\Sigma_c^+(2800)]=2792^{+14}_{-5}$ MeV/c$^2$ reported by the PDG based on Belle's measurements~\cite{Mizuk:2004yu}. Additionally, the two lowest states that we extract for the $\Sigma_c(\frac{1}{2}^-)$ with masses $M_1 = 2814 \pm 20$ MeV/c$^2$ and $M_2 = 2854 \pm 17$ MeV/c$^2$, might be candidates for yet unobserved $\Sigma_c$ excitations. Note that the three extracted negative parity $\Sigma_c$ states are well above their respective two-particle thresholds so that the two-particle contribution to the signals should be suppressed.

				\paragraph{Excited $\Xi_c$ and $\Xi_c'$ states:} 
				The experimental spectrum of the $\Xi_c$ and $\Xi_c'$ channels consists first of the respective $J^P = 1/2^{+}$ ground states, and the first $\Xi_c(\frac{1}{2}^{-})$ excited state, which are all experimentally well established and which we reproduce well in our work. The energy levels above the lowest three are less well established, both experimentally and theoretically. Above 2.9 GeV/c$^2$, the PDG reports the five states $\Xi_c(2930)$, $\Xi_c(2970)$, $\Xi_c(3055)$, $\Xi_c(3080)$ and $\Xi_c(3123)$, for none of which the spin and parity quantum numbers have been measured. Very recently, the spectrum of these states has received an update by a new measurement of the LHCb Collaboration \cite{Aaij:2020yyt} in the $\Lambda_c^+ K^-$ channel. According to this measurement, the $\Xi_c(2930)$ (observed earlier by the Belle~\cite{Li:2017uvv} and the BABAR~\cite{Aubert:2007eb} Collaborations in the same channel) should be considered to be a previously unresolved combination of two independent states $\Xi_c(2923)$ and $\Xi_c(2939)$. The third observed state in Ref.~\cite{Aaij:2020yyt}, $\Xi_c(2965)$, corresponds either to the already seen $\Xi_c(2970)$, or is another entirely new resonance.
				
				Let us discuss potential interpretations of our findings with regard to this rather rich experimental spectrum. We find two negative parity spin-$1/2$ $\Xi_c'$ states in the vicinity of the lowest three (or four) states above 2.9 GeV/c$^2$, $\Xi_c(2923)$, $\Xi_c(2939)$, $\Xi_c(2965)$ and potentially $\Xi_c(2970)$, which suggests that such quantum numbers can be assigned to at least two of these states. While our numerical results are not precise enough to draw any firm conclusions, our obtained spectrum is most naturally interpreted as either $\Xi_c(2923)$ or $\Xi_c(2939)$ and similarly $\Xi_c(2965)$ or $\Xi_c(2970)$ being a $\Xi_c'(\frac{1}{2}^{-})$ state. 
				
				The already known $\Xi_c(2970)$ state has been observed in the $\Lambda_c \overline{K} \pi$ channel -- also proceeding approximately half of the time via the intermediate $\Sigma_c(2455) \overline{K}$ channel -- and in the $\Xi_c^\prime \pi$, and $\Xi_c(2645) \pi$ channels by the Belle~\cite{Chistov:2006zj,Lesiak:2008wz,Yelton:2016fqw} and BABAR~\cite{Aubert:2007dt} Collaborations. These decay channels imply several possible quantum numbers, $J^P=(1/2^\pm,3/2^\pm,5/2^\pm)$, for this state, which is not in contradiction with the above potential assignment. 
				
				For the energy levels above 3.0 GeV/c$^2$, we obtain two states in the region of the states $\Xi_c(3055)$, $\Xi_c(3080)$ and $\Xi_c(3123)$, one $\Xi_c(\frac{1}{2}^{-})$ and one $\Xi_c'(\frac{1}{2}^{+})$ state, respectively. Again, the uncertainties of the numerical results are too large for definite assignments, but point to the possibility that one of the three measured states is either a $\Xi_c(\frac{1}{2}^{-})$ and a  $\Xi_c'(\frac{1}{2}^{+})$ state.
				
				The $\Xi_c(3055)$ was observed by the Belle and the BABAR Collaborations in the $\Sigma_c \overline{K}$ channel~\cite{Kato:2013ynr,Aubert:2007dt} and in the $\Lambda D$ channel only by the Belle Collaboration~\cite{Kato:2016hca}. Masses reported by the Belle Collaboration are $M[\Xi_c^0(3055)] = 3059.0 \pm 1.1$ MeV/c$^2$ and $M[\Xi_c^+(3055)] = 3055.8 \pm 0.6$ MeV/c$^2$, which are close to our second $\Xi_c(\frac{1}{2}^-)$ which lies above all the relevant lattice thresholds and the physical $\Lambda D$ threshold. 
				
				Finally, the $\Xi_c(3080)$ was reported by the Belle Collaboration~\cite{Kato:2016hca} in the $\Sigma_c \overline{K}$, $\Sigma_c^\ast \overline{K}$, and $\Lambda D$ channels and by the BABAR Collaboration~\cite{Aubert:2007dt} in the $\Lambda_c \overline{K} \pi$ channel via the $\Sigma_c(2455) \overline{K}$ channel. Similar to the $\Xi_c(2970)$ case, these decay channels suggest several quantum numbers, such as $J^P = (1/2^{\pm},3/2^{\pm},5/2^{\pm})$. Our second $\Xi_c^\prime(\frac{1}{2}^+)$ state appears to be the most probable candidate for this resonance.

				\paragraph{Excited $\Omega_c$ states:} The five new excited $\Omega_c^0$ states reported by the LHCb Collaboration \cite{PhysRevLett.118.182001} were seen in the $\Xi_c \overline{K}$ channel. One would hence naively expect these states to have negative parity. A first dedicated lattice QCD calculation has confirmed this expectation by assigning negative parity to these states~\cite{PhysRevLett.119.042001}, with total spin ranging from $J=1/2$ to $5/2$. The two $\Omega_c(\frac{1}{2}^-)$ states and the lowest-lying $\Omega_c^\ast(\frac{3}{2}^-)$ state that we extract lie in the vicinity of these excited $\Omega_c$ baryons observed by the LHCb Collaboration. The pattern depicted in \Cref{fig:threshold_comp} matches that of the experimental spectrum where there are two states closer to the $\Xi_c^\prime \overline{K}$ thresholds and one coinciding with the $\Xi_c^\prime \overline{K}$. The second $\Omega_c(\frac{1}{2}^-)$ and the lowest-lying $\Omega_c^\ast(\frac{3}{2}^-)$ states are close to the $\Xi_c' \overline{K}$ threshold. The statistical error of the $M_1[\Omega_c^\ast(\frac{3}{2}^-)]$ state spans most of the energy region of the LHCb states. It therefore at this stage is rather futile to draw any definite conclusions.   

				We should reiterate that since we only employ local three-quark operators, we are limited in our ability to resolve all molecular, radial or orbital excitation modes of the higher lying states. Our results should hence be considered as indicative in identifying potential compact three-quark states among the experimentally observed energy levels in the $\Xi_c$ and the $\Omega_c$ channels. Conversely, the levels that we are not able to reproduce, could be candidates for molecular or orbitally excited states. It is however at present too early to assign definite quantum numbers without a through scattering state analysis since some of our negative parity states lie close to the thresholds. 

\begin{figure*}
					\centering
					\includegraphics[width=.85\textwidth, trim={25mm 12.5mm 35mm 20mm},clip]{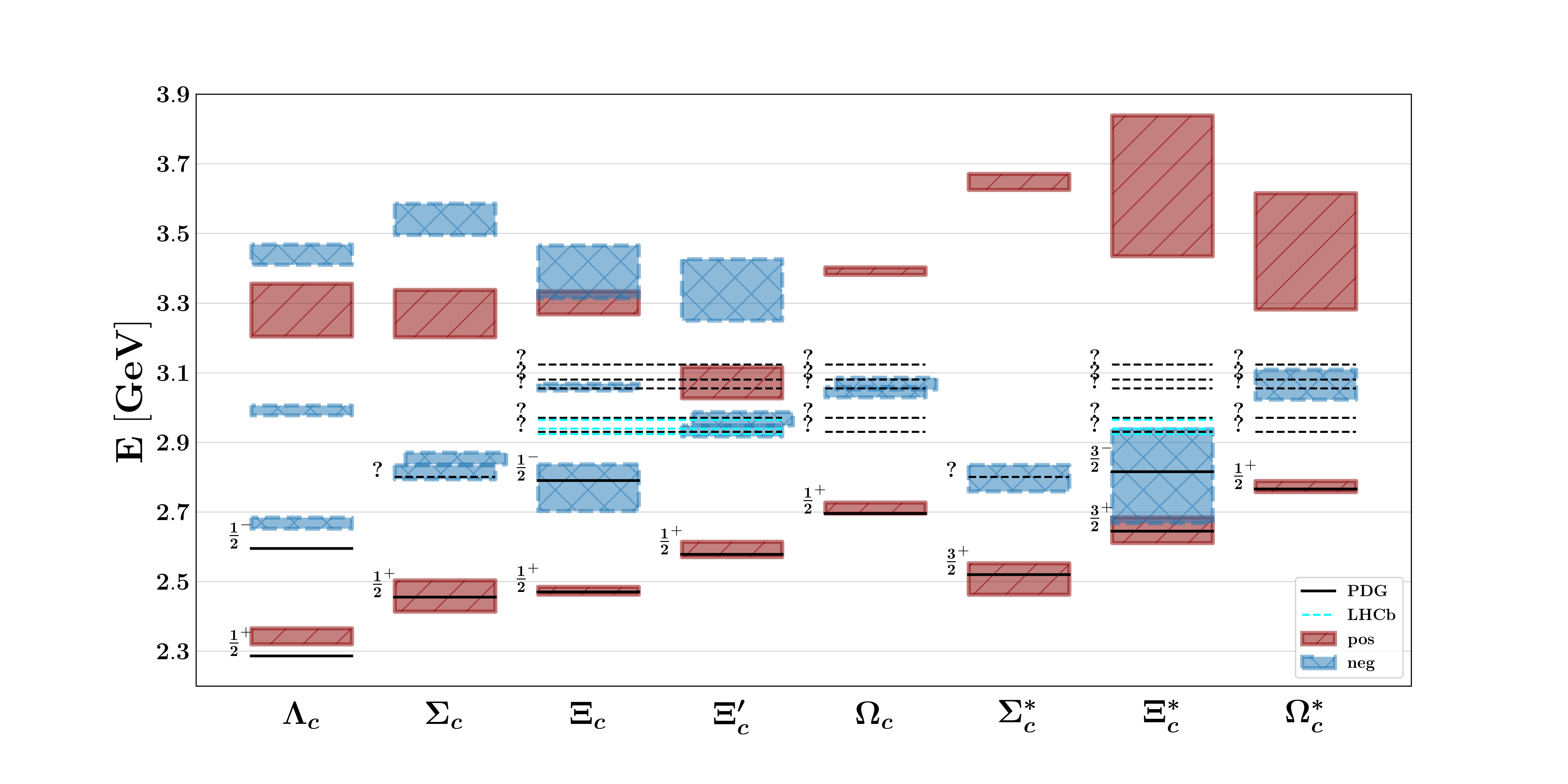}
					\caption{ \label{fig:finres} Our results from \Cref{tab:barmass} laid over related experimental results. Boxes indicate the statistical uncertainties. Close by states are shifted for clarity. All black and cyan lines are experimental results, solid (dashed) for the states with (un)determined quantum numbers. Recent LHCb results \cite{Aaij:2020yyt} (cyan dashed) in the $\Xi_c^{(\prime,\ast)}$ channels are included as well.}
				\end{figure*}

				The values in \Cref{tab:barmass} are illustrated in \Cref{fig:finres} together with the relevant experimental results. The latest $\Xi_c$ results from the LHCb Collaboration are shown as well. The similarities between the $\Lambda_c$ and $\Xi_c$, and $\Sigma_c$, $\Xi_c^\prime$ and $\Omega_c$ are evident as expected from their flavor structures.

				\paragraph{Interpretation from a quark model perspective}
				The quark model (QM) has has been useful in giving a pictorial and intuitive interpretation of the mass spectrum obtained by lattice QCD computations. The QM derives the energy and structure of a system by considering constituent valence quarks and their interactions. For the excited states, in particular, it can clarify what the essential degrees of freedom in a specific excitation are.

				For heavy-quark baryons, the heavy-quark spin symmetry plays an important role. As the coupling of a heavy quark to the magnetic component of gluons is suppressed by a $1/m_Q$ factor, the heavy quark spin is approximately conserved. For singly charmed baryons, this symmetry is manifested by the appearance of heavy-quark spin doublets, in which spin $(j-1/2, j+1/2)$ pair states approach each other 
				with increasing quark masses. Here, $j$ represents the total spin minus the heavy quark spin of the considered baryon.

				We will here briefly compare the present lattice QCD results with the QM predictions 
				and study how the essential excitation modes arise in the spectrum. 
				Quite remarkably, multiple features of the QM predictions are confirmed in the obtained lattice QCD spectrum of the charmed baryons.

				\begin{enumerate}
					\item Our lattice QCD results for the positive parity ``ground'' states agree completely with the QM assignments, {\it i.e.}, the spin, parity, isospin and flavor representation, and the mass orderings are consistent. The QM predictions for the splitting between the spin $1/2$ and $3/2$ states are also in quantitative agreement with the obtained lattice results.

					\item Among the positive parity ground states, $\Xi_c$ is most interesting, because it contains three different valence quarks, $c$, $s$, and $u/d$. In the QM, the total spin of $s$ and $u/d$ can take either $S=0$ ($\Xi_c$), or 1 ($\Xi_c'$). The existence of two low-lying positive parity states is indeed realized in lattice QCD as well as in experiment. In the QM, the distinction of $\Xi_c$ and $\Xi_c'$ is guaranteed by the flavor $SU(3)$ symmetry, while the $SU(3)$ breaking with $m_s\ne m_{u/d}$ will mix the two $\Xi_c$'s. The QM predicts, however, that the mixing is suppressed for the ground state due to the heavy quark spin symmetry, which is confirmed in our lattice QCD results.

					\item Low-lying negative parity singly charmed baryons are described in the QM as orbital $P$-wave excitations. They are categorized in two classes, $\lambda$-mode and $\rho$-mode~\cite{PhysRevD.20.768,Yoshida:2015tia}. The $\lambda$-mode is characterized by the $P$-wave excitation between the charm quark and the center of mass of the light quarks, while the $\rho$-mode is given by the excitation between the light quarks. The QM predicts that the $\lambda$ modes are lighter than the $\rho$-modes for singly heavy baryons.

					The QM spectrum depends on the flavor structure: For the flavor anti-triplet $\Lambda_c$ and $\Xi_c$, we find a set of ($1/2^-$, $3/2^-$) states in the $\lambda$-mode, and ($1/2^-$), ($1/2^-$, $3/2^-$) and ($3/2^-$, $5/2^-$) states in the $\rho$-mode. Thus, among the three $1/2^-$ states, the QM predicts that one $\lambda$-mode state is lighter than the other two. This structure is indeed seen in the $\Lambda_c$ and $\Xi_c$ spectrum given in \Cref{tab:barmass} and \Cref{fig:finres}. The next $1/2^-$ state is about 300 MeV higher, which can be regarded as the mass splitting between the $\lambda$- and $\rho$-mode states.

					On the other hand, the flavor {\bf 6} baryons, $\Sigma_c$, $\Xi_c'$ and $\Omega_c$, have two $\lambda$-mode $1/2^-$ states, one of them being accompanied by a $3/2^-$ state. In terms of the heavy-quark spin symmetry,  we have a ($1/2^-$, $3/2^-$) spin doublet and an isolated singlet $1/2^-$. The lower two $\lambda$-mode states come close in energy, but can be distinguished by the total angular momentum of the light-quark system. 
					Thus we expect two $1/2^-$ and one $3/2^-$ states as the lowest negative parity excitations for $\Sigma_c$, $\Xi_c'$ and $\Omega_c$. One sees that, indeed, these three states turn out to be almost degenerate in the lattice QCD spectrum of these channels in \Cref{tab:barmass} and \Cref{fig:finres}. Other states are much higher in energy, which again confirms the predicted QM assignments.
				\end{enumerate}
				In all, the low-lying spectra of both the positive and negative parity charmed baryons confirm the effectiveness of the QM in assigning the quantum numbers and symmetry properties of heavy baryons.

\begin{figure*}[htb]
					\centering
					\includegraphics[width=.49\textwidth]{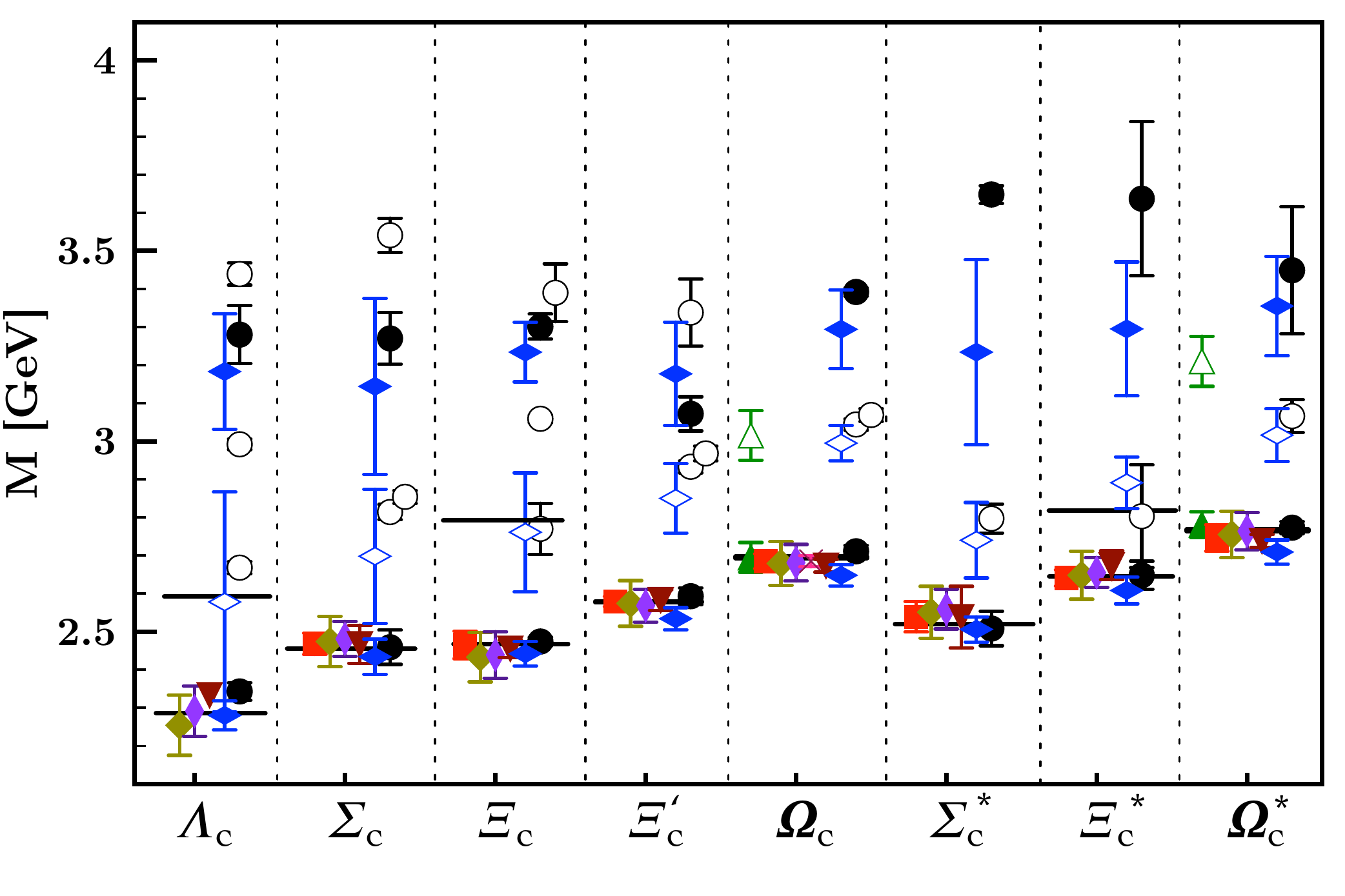} \hfill
					\includegraphics[width=.49\textwidth]{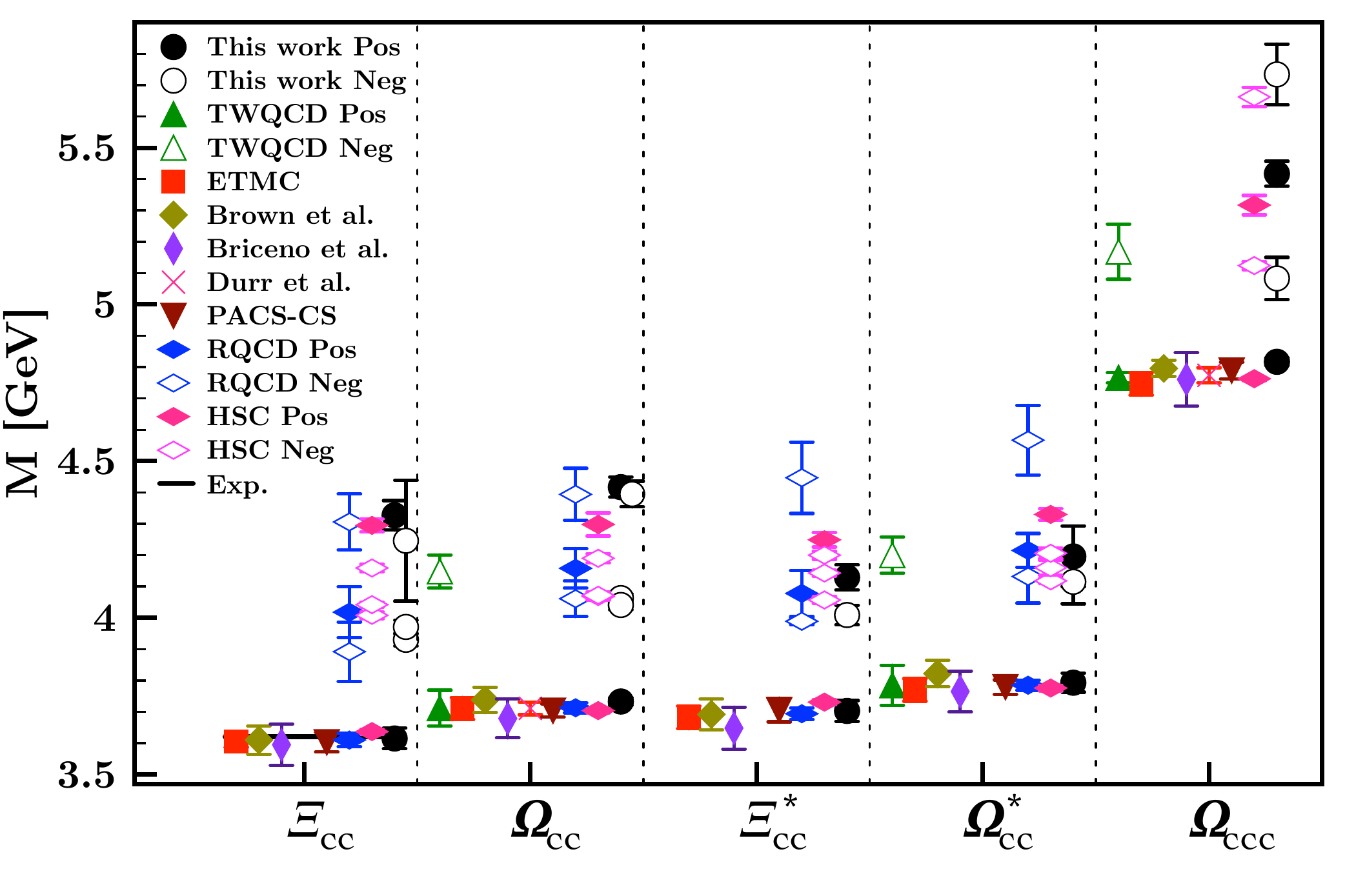}
					\caption{ \label{fig:comp} Our results in comparison with the determinations of the ETMC~\cite{Alexandrou:2017xwd}, D\"{u}rr \emph{et al.}~\cite{Durr:2012dw}, Brown \emph{et al.}~\cite{Brown:2014ena}, PACS-CS~\cite{Namekawa:2013vu}, TWQCD~\cite{Chen:2017kxr}, Brice\~no \emph{et al.}~\cite{Briceno:2012wt}, RQCD~\cite{PhysRevD.92.034504}, and HSC~\cite{Padmanath:2013zfa,Padmanath:2015jea}. Note that the lowest two data points of the HSC for the $\Omega_{cc}(\frac{1}{2}^-)$ baryon are almost on top of each other. Error bars are smaller than the symbols for some points. Only the lowest-lying experimental values are shown. }
				\end{figure*}

				\paragraph{Comparison to other lattice results:}
				We compare our results to other lattice determinations and experimental values in \Cref{fig:comp}. Our positive parity ground states are in good agreement with the experimental results and the calculations of the other lattice groups with the exception of the $\Lambda_c$, which is overestimated in our work. Taken altogether, this is a good indication that we are close to the physical point. The first excited positive parity states also mostly agree with the predictions of the HSC~\cite{Padmanath:2013zfa,Padmanath:2015jea} and the RQCD Collaboration~\cite{PhysRevD.92.034504}. For negative parity, there are notable differences between our and RQCD's results, especially for the doubly-charmed baryons. For the excited states of the $\Xi_{cc}$ and $\Omega_{cc}$, there are discrepancies between our extracted spectrum and that of RQCD, while our results are similar to those obtained by the HSC~\cite{Padmanath:2015jea}. Although we do not show the corresponding HSC spectrum in \Cref{fig:comp}, the pattern they extract in their preliminary studies for the negative parity spin-$1/2$ singly-charmed baryons~\cite{Padmanath:2015bra} is similar to our results as well. Such a qualitative agreement for the low-lying spectrum is quite encouraging since, in contrast to the HSC, which utilizes both local and non-local operators, we only use local operators.  

				\paragraph{Systematic uncertainties:} 
				Finally, we comment on possible systematic effects that have not been addressed in this work. As stated before, these particular PACS-CS configurations have $m_\pi L < 4$ which would suggest that there may be significant finite size corrections. Since we carry this work on a single volume we are unable to quantify such systematics. However, we have shown in Ref.~\cite{Can:2015exa} that the finite size effects are negligible for the ground state charmed-strange baryon systems which might provide an indication for the current study although a further investigation would be desirable to confirm our results. 

				Although we have inspected the scattering states for the negative parity channels, a thorough study based on a L{\"u}scher approach would be needed to fully quantify the contamination from these states. Additionally, strong decays of the positive parity states are not taken into account. The sole example for the ground state charmed baryons would be the $\Sigma_c \to \Lambda_c \pi$ decay. However we note that the ground state $\Sigma_c$ is not a resonant state but a bound state in our lattice setup since there is not enough phase space for the decay to occur with respect our extracted $M_{\Sigma_c} - M_{\Lambda_c} \sim 116$ MeV splitting. Excited positive parity signals on the other hand lie too high and it would be irrelevant at this stage to consider them. 

				The relativistic heavy quark action we employ removes the leading order cutoff effects of order $\mathcal{O}\left( (m_q a)^n \right)$ and reduces them to $\mathcal{O}\left( (a\Lambda_{\text{QCD}})^2 \right)$ by a proper tuning of the action parameters. However, in order to fully remove these effects a continuum extrapolation is necessary. Since we extract the spectrum on a single lattice spacing, no continuum extrapolation is performed and such an effect is essentially still present although it can be expected to be negligible compared to the present statistical uncertainties.

	\section{Summary and Conclusions}\label{sec:sc}
		We have calculated the ground and the first few excited states of the charmed baryons on $2+1$-flavor gauge configurations with a pion mass of $\sim 156$ MeV/c$^2$. The charm quark is treated relativistically by employing a relativistic heavy-quark action to remove $O(a m_Q)$ discretization errors. The states are extracted via a variational approach over a set of interpolating fields with different Dirac structures and quark-field smearings. By performing separate variational analyses with multiple subsets of the operator basis, we have studied the Dirac-structure and smearing dependence of the excited states. Our results indicate that the excited-state signals are highly susceptible to the width of the quark smearing. Additionally, solutions of a variational analysis over a set of smeared operators with fixed Dirac structure couple dominantly to the operator that is smeared the widest within our employed smearing parameter range. These results highlight the importance of forming the variational basis from different Dirac structures since relying on smeared operators only might miss some parts of the spectrum. 

		In comparing the operator dependence of the extracted positive and negative parity states, we have extended the $SU(4)$ operator basis of the $\Xi_c$ baryons to include not only $\Lambda$-like, but also $N$-like operators. Both operators give consistent results for the positive parity case while there appears a difference for the negative parity states. We have also investigated the $\Xi_c$--$\Xi_c^\prime$ mixing by studying the cross-correlators of this system.  

		Our masses of the low-lying states agree well with the available experimental results and previous lattice determinations. Consequently, the hyperfine splittings and the mass differences between the positive and negative parity states are reproduced, which is a good check of the relativistic action we employ for the charm quark. Excited states in the positive parity channel lie 400 MeV to 1 GeV above the ground states depending on the quantum numbers. One or more negative parity states appear in between. This pattern is consonant with the QM expectations. Although we identify several states that are close to observed excited $\Sigma_c$, $\Xi_c$ and $\Omega_c$ baryons, mostly in the negative parity channels, some of the signals are in close proximity to the related two-particle thresholds. Without a thorough scattering state analysis with multiple volumes and two-particle operators, the contamination from the thresholds remain unidentified. 

		From a qualitative point of view, the spectrum we extract is similar to what has been reported by the Hadron Spectrum Collaboration (HSC). This is quite encouraging since the HSC employs a large operator basis including nonlocal operators. The qualitative agreement indicates the practicality of using local operators to probe the low-lying excitations, even though further work especially regarding the proper treatment of scattering states is
		still needed.   

	\acknowledgments
		K.U.C. thanks M. Padmanath for discussions and sharing the Hadron Spectrum Collaboration's doubly-charmed baryon results. The unquenched gauge configurations employed in our analysis were generated by the PACS-CS Collaboration~\cite{PhysRevD.79.034503}. We have downloaded the publicly available configurations via the ILDG/JLDG network~\cite{Amagasa:2015a, ildg_jldg}. This work is supported in part by The Scientific and Technological Research Council of Turkey (TUBITAK) under project number 114F261 and in part by KAKENHI under Contract Nos. JP18K13542, JP19H05159, JP20K03940 and JP20K03959. K.U.C is supported in part by the Special Postdoctoral Researcher (SPDR) program of RIKEN and in part by the Australian Research Council Grant DP190100297 during the course of this work. H.B. acknowledges financial support from the Scientific and Technological Research Council of Turkey (TUBITAK) BIDEB-2219 Postdoctoral Research Programme. P.G. is supported by the Leading Initiative for Excellent Young Researchers (LEADER) of the Japan Society for the Promotion of Science (JSPS).

	% \bibliography{barspec}

\end{document}